\documentclass[journal,comsoc]{IEEEtran}

\usepackage{times}
\usepackage[cmex10]{amsmath}
\usepackage{amssymb}
\usepackage{epsfig,verbatim}
\usepackage{algorithm}
\usepackage{algpseudocode}

\usepackage{graphicx,color,epsfig,rotating,subfigure}
\usepackage{amsfonts,amsmath,amssymb}
\usepackage{algorithm}
\usepackage{subfigure}
\usepackage{algpseudocode}

\newcommand\blfootnote[1]{%
  \begingroup
  \renewcommand\thefootnote{}\footnote{#1}%
  \addtocounter{footnote}{-1}%
  \endgroup
}

\long\def\comment#1{}

\newfont{\bbb}{msbm10 scaled 700}

\newfont{\bb}{msbm10 scaled 1100}

\newcommand{\bv}{{\bf b}}

\newcommand{\pv}{{\bf p}}

\newcommand{\sv}{{\bf s}}

\newcommand{\uv}{{\bf u}}

\newcommand{\xv}{{\bf x}}
\newcommand{\yv}{{\bf y}}

\newcommand{\onev}{{\bf 1}}

\newcommand{\Gm}{{\bf G}}

\newcommand{\Pm}{{\bf P}}

\newcommand{\Ac}{{\cal A}}
\newcommand{\Bc}{{\cal B}}
\newcommand{\Cc}{{\cal C}}
\newcommand{\Dc}{{\cal D}}

\newcommand{\Ic}{{\cal I}}

\newcommand{\Lc}{{\cal L}}

\newcommand{\Sc}{{\cal S}}
\newcommand{\Tc}{{\cal T}}

\newcommand{\Wc}{{\cal W}}

\newcommand{\Xc}{{\cal X}}
\newcommand{\Yc}{{\cal Y}}

\newcommand{\eqdef}{\stackrel{\Delta}{=}}

\newtheorem{theorem}{Theorem}

\newtheorem{definition}{Definition}

\newtheorem{remark}{Remark}

\newtheorem{proposition}{Proposition}

\newtheorem{example}{Example}

\title{An Efficient Construction of Rate-Compatible Punctured Polar (RCPP) Codes Using\\ Hierarchical Puncturing}

\author{
               Song-Nam Hong,~\IEEEmembership{Member,~IEEE} and Min-Oh Jeong,~\IEEEmembership{Student Member,~IEEE} 
\thanks{The material in this paper was presented in part at the 2018 IEEE International Symposium on Information Theory.

The authors are with the Department of Electrical and Computer Engineering, Ajou University, Suwon, Korea (email: \{snhong, jmo0802\}@ajou.ac.kr).

This work was supported by the Future Combat System Network Technology Research Center program of Defense Acquisition Program Administration and Agency for Defense Development (UD160070BD).

}
}

\begin{document}

\maketitle

\date{}

\blfootnote{
}

\begin{abstract}
 In this paper, we present an efficient method to construct a good rate-compatible punctured polar (RCPP) code for incremental redundancy hybrid automatic repeat request (IR-HARQ) schemes. One of the major challenges on the construction of a RCPP code is to optimize a common information set which is good for all the (punctured) polar codes in the family. Unfortunately, there is no efficient way to solve the above problem. In the proposed construction, a common information set is simply optimized for the highest-rate code in the family and then it is updated to yield an effective information set for each other code, by keeping the condition that  information bits are unchanged during retransmissions. This is enabled by presenting a novel {\em hierarchical} (or {\em reciprocal}) puncturing and {\em information-copy} technique. Specifically, some information bits are copied to frozen-bit channels whose locations are carefully determined according to rate-compatible puncturing patterns. This yields an information-dependent frozen vector in the encoding part. Also, in the decoding part, the effective information sets are obtained by properly combining the common information set and the information-dependent frozen vector.  
 More importantly, the impact of ``unknown" frozen bits are avoided due to the special structure of the proposed hierarchical (or reciprocal) puncturing. Simulation results verify that the proposed RCPP code can yield a significant performance gain (about 2dB) over a benchmark RCPP code where both codes use the same rate-compatible puncturing patterns but the latter uses the conventional all-zero frozen vector. Therefore, the proposed method would be crucial to construct a good RCPP code efficiently.
\end{abstract}

\section{Introduction}

Polar codes, proposed by Arikan \cite{Arikan2009}, achieve the symmetric capacity of binary-input discrete memoryless channels (BI-DMCs) under a low-complexity successive cancellation (SC) decoder. The finite-length performances of polar codes can be enhanced by using a list decoder that enables polar codes to approach the performance of the optimal maximum-likelihood (ML) decoder \cite{TalVardy2011}. It was further shown in \cite{TalVardy2011} that a polar code  concatenated with a simple CRC can outperform well-optimized LDPC and Turbo codes especially for short lengths. Due to their good performance and low-complexity, polar codes are currently considered as channel codes in future wireless communication systems (i.e., 5G cellular systems).

Wireless broadband systems (e.g., 4G LTE and 5G) operate in the presence of time-varying channels, which requires flexible and adaptive transmission techniques. In these systems,  incremental redundancy  hybrid automatic repeat request (IR-HARQ)  schemes are widely employed in which  parity bits for retransmission are chosen in an {\em incremental} fashion according to a certain rate requirement. They are enabled by the use of a  {\em rate-compatible} (RC) code which consists of a family of codes to support various rates. For the RC code, it should be guaranteed that the set of parity bits of a higher-rate code is a subset of the set of parity bits of a lower-rate code, which is referred to as {\em rate-compatibility constraint}. This is able to allow the receiver that fails to decode at a particular rate, to request only additional parity bits from the transmitter. For this reason, there have been  extensive researches on the construction of RC Turbo and RC LDPC codes (see \cite{Rowitch2000, El-Khamy2009, Tsung-Yi2015} and the references therein).

Very recently, a capacity-achieving RC polar code, named parallel concatenated polar (PCP) code, was presented in  \cite{Hong-IT} for IR-HARQ schemes. In this method, a capacity-achieving (punctured) polar code can be used for every transmission by satisfying the rate-compatibility constraint. In \cite{IF}, a similar method was independently presented by the name of incremental freezing. Although both methods can achieve the optimal performance for sufficiently large lengths, it does not directly imply that they yield attractive performances for practical lengths. In particular when incremental rate is small, they cannot perform well as the length of constituent (punctured) polar code is too small.  An alternative approach to design a RC polar code by performing puncturing successively from a mother polar code, which is referred to as RC punctured polar (RCPP) code. On the construction of a RCPP code, it is required to jointly optimize rate-compatible puncturing patterns and a common information set (which is good for all the codes in the family). Unfortunately, this optimization is not tractable due to its extensive complexity.  Instead, numerous heuristic methods were presented to generate a puncturing pattern and an associated information set (see \cite{Eslami2011},  \cite{Chen2013}, \cite{WangLiu2014}, \cite{Shin2013} and the references therein). In \cite{Niu}, a practical puncturing pattern, named quasi-uniform puncturing (QUP), was presented and shown to yield an attractive performance. However, its extension to a rate-compatible puncturing is not straightforward because of the design of a good common information set. In \cite{Miloslavskaya2015}, an efficient search algorithm to jointly optimize a puncturing pattern and an information set was developed and shown to outperform LDPC codes. It is remarkable that, in the existing methods \cite{WangLiu2014}-\cite{Miloslavskaya2015}, an optimized information set according to each puncturing pattern was employed, i.e., each code in the family can use its own optimized information set. These approaches cannot be applied to IR-HARQ schemes since information bits, in this case, should be kept during retransmissions (i.e., a common information should be used for all the codes in the family). Therefore, it is still open problem to construct a good RCPP code for IR-HARQ schemes.


In this paper, we present an efficient method to construct a good RCPP code for IR-HARQ schemes. In fact, there are good rate-compatible puncturing patterns like QUP in \cite{Niu} as long as an optimized information set can be used for each puncturing pattern. As pointed out before, it is not possible since information sets, in IR-HARQ schemes, should be unchanged during retransmissions. Motivated by this, we propose a novel technique to produce {\em effective} (or virtual) information sets from the common one, each of which is the optimized information set for the corresponding code in the family.

Our contributions are summarized as follows.

\begin{itemize}
\item We present a novel {\em hierarchical puncturing} which has the special property such that ``unknown" frozen bits can be allocated to some frozen-bit channels (carefully chosen according to a puncturing pattern), without affecting the performance. 
Moreover, we derive a class of hierarchical puncturing patterns which satisfy the rate-compatible constraint (e.g., QUP).

\item We then propose an {\em information-copy} technique which repeats some information bits to frozen-bit channels as well as information-bit channels, which can yield an {\em information-dependent} frozen vector. Here, the locations of non-zero frozen-bit channels are determined as a function of rate-compatible puncturing patterns and the associated optimized information sets.

\item Leveraging hierarchical puncturing and information-dependent frozen vector, we develop a systematic method to construct a good RCPP code for IR-HARQ scheme. In the proposed method, a common information set is first optimized for the highest-rate code in the family and then, it is updated to produce {\em effective} information sets which are optimized for the other codes in the family, by properly exploiting the common information set and information-dependent frozen vector. Thanks to the property of hierarchical puncturing, moreover, the impact of ``unknown"  (information-dependent) frozen bits are completely avoided.

\item Finally, simulation results demonstrate that the proposed RCPP code yields a considerable performance gain (about 2dB) over a benchmark RCPP code for IR-HARQ scheme, where  both codes use the same QUP. The main difference of the both methods is that the former uses an information-dependent (non-zero) frozen vector while the latter uses a conventional all-zero frozen vector. Therefore, the proposed construction method would be crucial to construct a good RCPP code for IR-HARQ scheme.

\end{itemize}


The outline of this paper is as follows. In Section~\ref{sec:pre}, we provide some useful notations and definitions to be used throughout the paper. In Section~\ref{sec:HPP}, We present novel reciprocal and hierarchical puncturing and derive their key properties for the construction of the proposed RCPP code. In Section~\ref{sec:RC}, leveraging the proposed puncturing and information-copy technique, we propose an efficient method to construct a good RCPP code. In Section~\ref{sec:SIM}, simulation results are provided to verify the superiority of the proposed RCPP code. Section~\ref{sec:conc} concludes the paper.

\section{Preliminaries}\label{sec:pre}

In this section we provide some useful notations and definitions that will be used in the sequel.

\subsection{Notation}

A polar code of length $N=2^n$ is considered, in which the synthesized (or polarized)  channels are indexed by $0,1,\ldots,N-1$. Let $\Ac\subseteq \{0,...,N-1\}$ denote the {\em information-bit set} which contains all the indices of the synthesized channels to carry information bits. Accordingly, its complement set $\Ac^c$ represents the {\em frozen-bit set} that contains all the indices of frozen-bit channels.  Let $\Gm_{{\rm N}} = \Gm_{2}^{\otimes n}$ be the rate-one generator matrix of all length-$N$ polar codes, where $\Gm_{2}$ denotes the  2-by-2 Arikan's Kernel \cite{Arikan2009} as
\begin{equation}\label{eq:G2}
\Gm_{2} \eqdef \left[\begin{array}{cc}
         1 & 0\\
         1 & 1\\
  \end{array}\right].
\end{equation} A length-$N$ polar code is specified with $\Gm_{N}$ and its information-bit set $\Ac$.

Given index subsets  $\Bc, \Dc \subseteq \{0,...,N-1\}$, let $\Gm_{{\rm N}}(\Bc,\Dc)$ denote the submatrix of $\Gm_{{\rm N}}$ obtained by selecting the rows and columns whose indices belong to $\Bc$ and $\Dc$, respectively. Define a function $g(\ell): \{0,...,N-1\} \rightarrow \{0,1\}^{n}$ which maps $\ell$ onto a {\em binary} expansion as
\begin{equation}\label{eq:g}
g(\ell)=(b^{\ell}_{n},\ldots,b^{\ell}_{1}),
\end{equation}such that $\ell=\sum_{i=1}^n b_i^{\ell} 2^{i-1}$. We let $w_{\rm H}(\bv)$ denote the number of non-zero elements in a vector $\bv$ (called the Hamming weight). Given $\uv_{{\rm N}}=(u_0,...,u_{N-1})$ and $\Ac\subset\{0,...,N-1\}$, we write $\uv_{\Ac}$ to represent the subvector $(u_i:i\in\Ac)$. 

\begin{figure}
\centerline{\includegraphics[width=9cm]{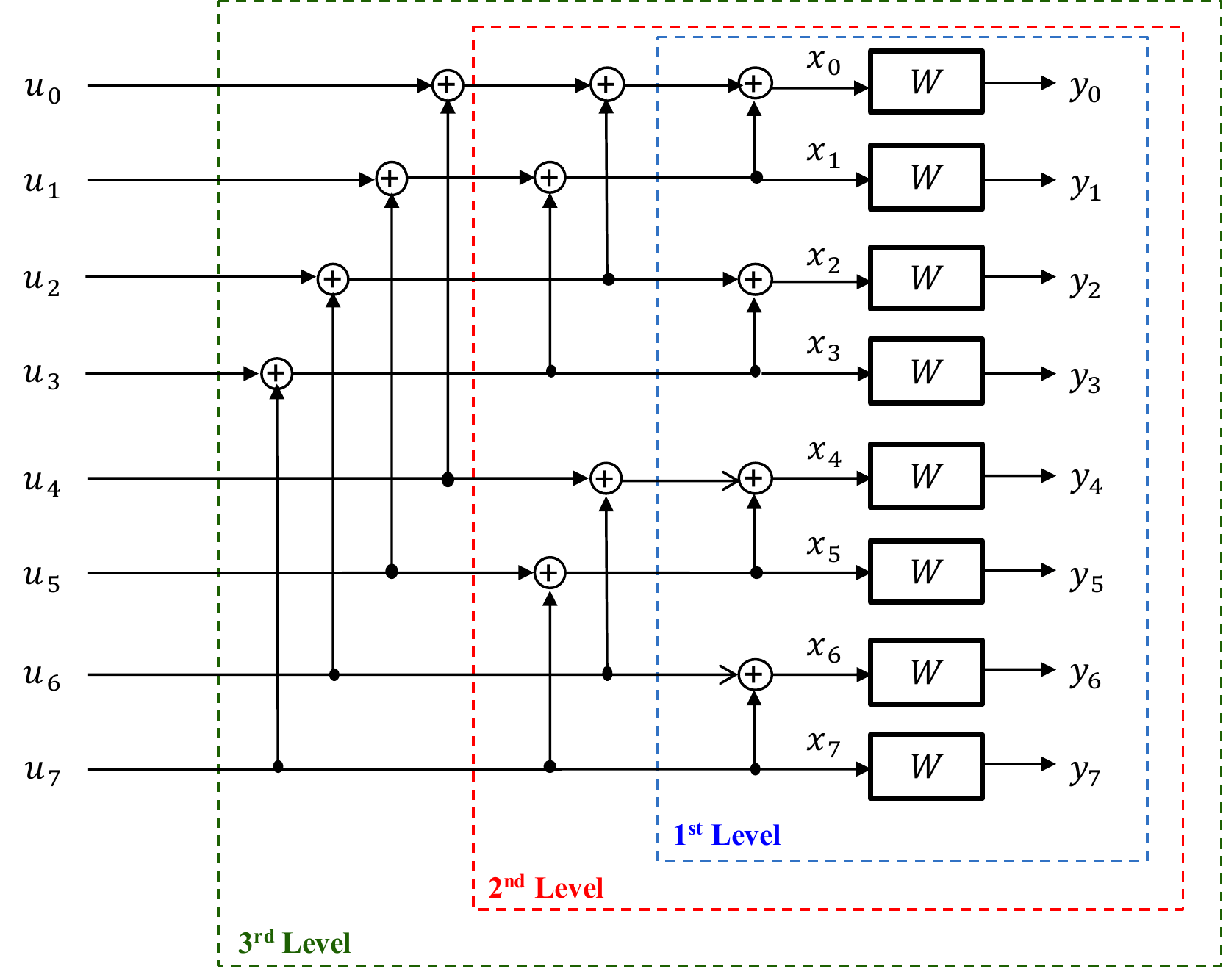}}
\caption{The length-$8$ polar code with input vector $\uv_8=(u_0,...,u_7)$ and output vector $\xv_8=(x_0,...,x_7)$. Also, given BI-DMC $\Wc:\Xc \rightarrow \Yc$, $x_i$ and $y_i$ represent the respectively channel input and output, where $x_{i} \in \Xc$ and $y_i \in \Yc$ for $i=0,...,7$.}
\label{fig:PC}
\end{figure}

\subsection{Punctured Polar Codes}\label{subsec:PPC}

In this section, we formally define a punctured polar code and its synthesized channels. Let $\uv_{\rm N}=(u_0,...,u_{{\rm N}-1})$ and  $\xv_{\rm N}=(x_0,...,x_{{\rm N}-1})$ be respectively the input and output vectors of a length-$N$ polar code. As shown in Fig.~\ref{fig:PC}, the encoding of the polar code is given by
\begin{equation}
\uv_{\rm N}\Gm_{\rm N} = \xv_{\rm N}.\label{eq:PC_ENC}
\end{equation} For the ease of expression, it is assumed that the bit-reverse permutation, denoted by $\psi(\cdot)$, is applied to the decoding part, instead of the encoding part as in \cite{Arikan2009}. Namely, the SC decoding successively decodes the $\hat{u}_{\psi(i)}$ for $i=0,...,N-1$ in that order. In the example of $N=8$, the SC decoding order is given as
 \begin{equation}
 \hat{u}_0 \rightarrow \hat{u}_4 \rightarrow \hat{u}_2 \rightarrow \hat{u}_6 \rightarrow \hat{u}_1 \rightarrow \hat{u}_5 \rightarrow \hat{u}_3 \rightarrow \hat{u}_7.
 \end{equation}

\begin{figure}
\centerline{\includegraphics[width=9cm]{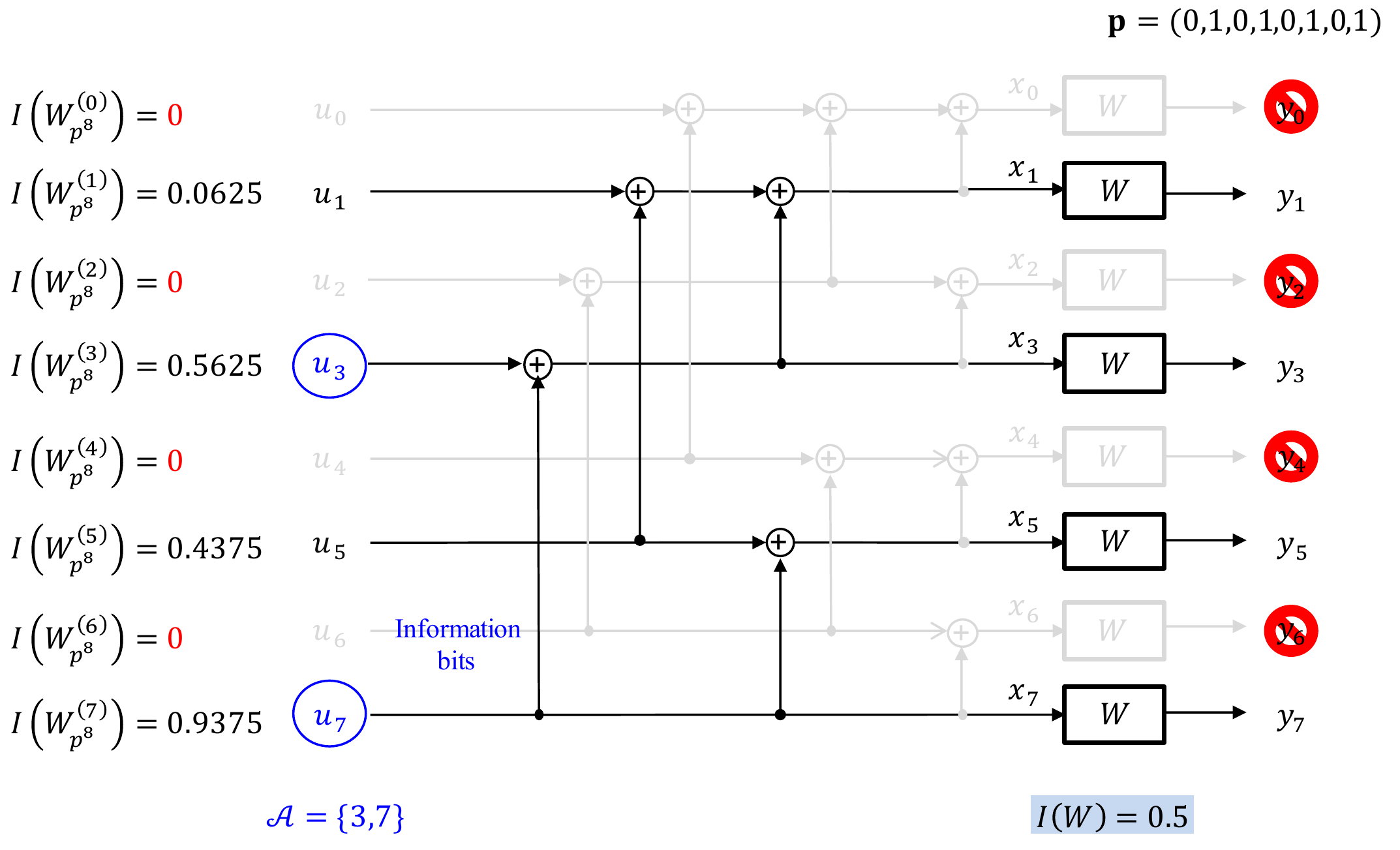}}
\caption{The punctured polar code of $N_p = 4$ where the information set is $\Ac=\{3,7\}$ and the puncturing pattern is $\pv = (0,1,0,1,0,1,0,1)$. Also, the input and output vectors are $\uv_{4}=(u_1,u_3,u_5,u_7)$ and $\xv_{4}=(x_1,x_3,x_5,x_7)$,  respectively.}
\label{fig:R_PC}
\end{figure}

A punctured polar code of a length $N_p < N$ is constructed by eliminating $N-N_p$ coded bits. Formally, it is described by the ``mother" polar code of the length $N$ and a binary vector (called the puncturing pattern)  $\pv_{\rm N}=(p_0,\ldots,p_{N-1}) \in \{0,1\}^{N}$ such that $w_{\rm H}(\pv_{\rm N}) = N_p$. Here,  $p_i = 0$ indicates that the $i$-th coded bit (e.g., $x_i$) is punctured and thus not transmitted. For simplicity, we will drop the index $N$ in 
$\pv_{\rm N}$ as long as it is identified in the context. Given $\pv$, we define the {\em zero-location set} which contains the locations of punctured bits  as 
 \begin{equation}
\Bc_{\pv}\eqdef \left\{i \in \{0,...,N-1\} : p_i = 0\right\}.\label{eq:zp_set}
\end{equation} In this paper, we specify a puncturing pattern using either a binary vector $\pv$ or a zero-location set $\Bc_{\pv}$. The corresponding unpunctured coded bits are 
denoted by $\xv_{{\rm N_P}} =(x_i : i \in \Bc_{\pv}^{c})$ and accordingly, the $N_p$ channel observations are denoted by $\yv_{{\rm N_P}} =(y_i : i \in \Bc_{\pv}^{c})$. The notion of synthesized channels in polar codes can be extended into punctured polar codes in a straightforwardly manner as follows. Given a  BI-DMC $W: {\cal X} \rightarrow {\cal Y}$, where $\Xc=\{0,1\}$ and $\Yc$ denote the respectively  input and output alphabets, a length-$N$ polar code, and an associated puncturing pattern $\pv$, the transition probability of the $i$-th synthesized channel of the resulting punctured polar code is defined as
\begin{align}\label{BitChannelPuncturedPolarCode}
& W^{(i)}(\yv_{{\rm N_{p}}},\uv_{i-1},\pv|u_i) \nonumber\\
&\;\;\;\;\;= \frac{1}{2^{N-1}} \sum_{u_{i+1},...,u_{N}} \sum_{\yv_{\rm N} \in \pi_{\pv}\left(\{\yv_{{\rm N_{p}}}\}\right)} W^{N}(\yv_{\rm N}| \uv_{\rm N}\Gm_{\rm N}),
\end{align}
where $\pi_{\pv}(S) \triangleq \{\yv_{\rm N} \in {\cal Y}^{N}: (y_i: i \in \Bc_{\pv}^c) \in S\}$. Also, the channel transition probabilities are given by
\begin{equation}
W^{N}(\yv_{\rm N}|\xv_{\rm N}) = \prod_{i \in [1:N]} W(y_i|x_i),\label{eq:tran}
\end{equation} where $W(\cdot|\cdot)$ denotes the channel transition probability of the underlying BI-DMC. Throughout the paper, we let  $W_{\pv}^{(i)}$ denote the $i$-th synthesized channel with the transition probability in (\ref{BitChannelPuncturedPolarCode}), and let $I(W_{\pv}^{(i)})$  denote the corresponding symmetric capacity. Then, an information set $\Ac$ of a punctured polar code is constructed by taking the indices of the $|\Ac|$ highest capacities as
\begin{equation}\label{eq:A-set}
\Ac=\{\ell_1,...\ell_{\Ac}\},
\end{equation} where $I(W_{\pv}^{(\ell_1)}) \geq I(W_{\pv}^{(\ell_2)}) \geq \cdots I(W_{\pv}^{(\ell_{|\Ac|})})$. Also, a punctured polar code is defined as  $\Cc(\Gm_{N}, \uv_{\Ac^c},\Ac,\pv)$ where $\uv_{\Ac^c}$ represents a frozen-bit vector. The encoding system of a punctured polar code is depicted in Fig.~\ref{fig:R_PC}.

\vspace{0.2cm}
\begin{remark} \label{remark:information-set} As seen in (\ref{eq:A-set}), an information set of a punctured polar code can be easily determined as long as $\{I(W_{\pv}^{(i)})\}$ (e.g., the capacities of the synthesized channels when a puncturing $\pv$ is applied), are computed. Even if they are well-defined mathematically, their computational complexities are generally unmanageable. Instead, the reliabilities of the synthesized channels can be efficiently computed by several techniques as density evolution under a Gaussian approximation (DE/GA), tracking the mean value, Battacharyya parameter of mutual information of Gaussian L-densities \cite{Sarkis}. Also, in 3GPP standardization, the simplest approach using a predetermined ordering sequence is presented \cite{3GPP}. In this sense, we use mutual information $\{I(W_{\pv}^{(i)})\}$ to explain the main idea and to describe the proposed algorithms while, in simulations, we use the DE/GA method.\flushright$\blacksquare$
\end{remark}

Given a puncturing pattern $\pv$, we define the {\em zero-capacity set} which contains the zero-capacity synthesized channels as
\begin{equation}
\Dc_{\pv}^{W}\eqdef \{ i \in \{0,...,N-1\}: I(W_{\pv}^{(i)}) = 0\}.\label{eq:zc_set}
\end{equation} Especially when $W$ is a perfect channel (i.e., noiseless deterministic channel), the above set is denoted by 
$\Dc_{\pv}$. Due to the nesting property of synthesized channels \cite{Korada}, we have
\begin{equation}
\Dc_{\pv} \subseteq \Dc_{\pv}^{W}.
\end{equation}  Furthermore, it was shown in \cite{Hong-TCOM} that 
\begin{equation}
|\Bc_{\pv}|=|\Dc_{\pv}|=N-w_{\rm H}(\pv),
\end{equation} namely, the number of zero-capacity synthesized channels is equal to that of punctured coded bits. Therefore, all the synthesized channels whose indices belong to $\Dc_{\pv}$ should be frozen-bit channels. Also, given $\{I(W_{\pv}^{(i)})): i=0,...,N-1\}$, we let 
\begin{align}
\Lc&=\mbox{max-ind}^{(t)} \{I(W_{\pv}^{(i)}): i=0,1...,N-1\}\\
\Sc&=\mbox{min-ind}^{(t)} \{I(W_{\pv}^{(i)}): i=0,1,...,N-1\},
\end{align} where $\Lc$ and $\Sc$ contain the indices corresponding to the $t$ largest and smallest values in 
$\{I(W_{\pv}^{(i)})\}$, respectively.

 \section{The Proposed Hierarchical Puncturing}\label{sec:HPP}
 
In this section, we propose a novel {\em hierarchical} puncturing which will be used as a key technology on the construction of the proposed RCPP codes. As noticed in Section~\ref{subsec:PPC}, all the synthesized channels associated with the zero-capacity set $\Dc_{\pv}$ should be frozen-bit channels, i.e.,  
\begin{equation}\label{eq:A-set2}
u_{i} = 0 \mbox{ for } i\in \Dc_{\pv} \Rightarrow \Ac \cap \Dc_{\pv} = \phi.
\end{equation} Otherwise, the corresponding punctured polar code surely leads to a frame (or block) error. It is required to identify $\Dc_{\pv}$ to construct a good information set $\Ac$. In \cite{Hong-TCOM}, it was shown that there exists a class of puncturing patterns to satisfy $\Dc_{\pv}=\Bc_{\pv}$, which are referred to as {\em reciprocal}. Moreover, their sufficient and necessary conditions are derived as follows.

\vspace{0.2cm}
\begin{theorem}[\cite{Hong-TCOM}]\label{thm:rec} A puncturing pattern $\pv$ is {\em reciprocal} if and only if the following properties are satisfied:
\begin{align*}
&\mbox{{\bf zero-inclusion: }} 0 \in \Bc_{\pv}\\
&\mbox{{\bf one-covering: }}  \mbox{For any } i \in \Bc_{\pv}, \mbox{ we have } j\in \Bc_{\pv} \mbox{ for all } j\\
&\;\;\;\;\;\;\;\;\;\;\;\;\;\;\;\;\;\;\;\;\;\;\;\;\;\;\;\;\;\;\;\;\;\;\;\;\;\;\;\;\;\;\;\;\;\;\;  \mbox{ such that } i \succeq_{1} j,
\end{align*} where $ i \succeq_{1} j$ implies that, if $b_{k}^{(j)} = 1$, then $b_{k}^{(i)} = 1$ for all $k=1,...,n$ and $ i \succeq_{1} 0$ for every $i>0$, and where $g(i) = (b_{n}^{(i)},...,b_{1}^{(i)})$ and $g(j) = (b_{n}^{(j)},...,b_{1}^{(j)})$.\flushright$\blacksquare$
\end{theorem}

For reciprocal puncturing patterns, the zero-capacity set $\Dc_{\pv}$ is straightforwardly identified from the puncturing pattern 
 $\pv$ (equivalently, $\Bc_{\pv}$), which makes it much easier to design a good information set $\Ac$ for the punctured polar code. In Theorem~\ref{thm:rec}, the one-covering property implies that if $p_7 = 0$ in a reciprocal puncturing pattern $\pv=(p_0,...,p_{15})$, then the following locations should be punctured: 
\begin{itemize}
\item weight-2 locations: $g^{-1}((0,0,1,1))$, $g^{-1}((0,1,0,1))$, and $g^{-1}((0,1,1,0))$;
\item weight-1 locations: $g^{-1}((0,0,0,1))$, $g^{-1}((0,0,1,0))$, and $g^{-1}((0,1,0,0))$.
\end{itemize}

\begin{figure}
\centerline{\includegraphics[width=9cm]{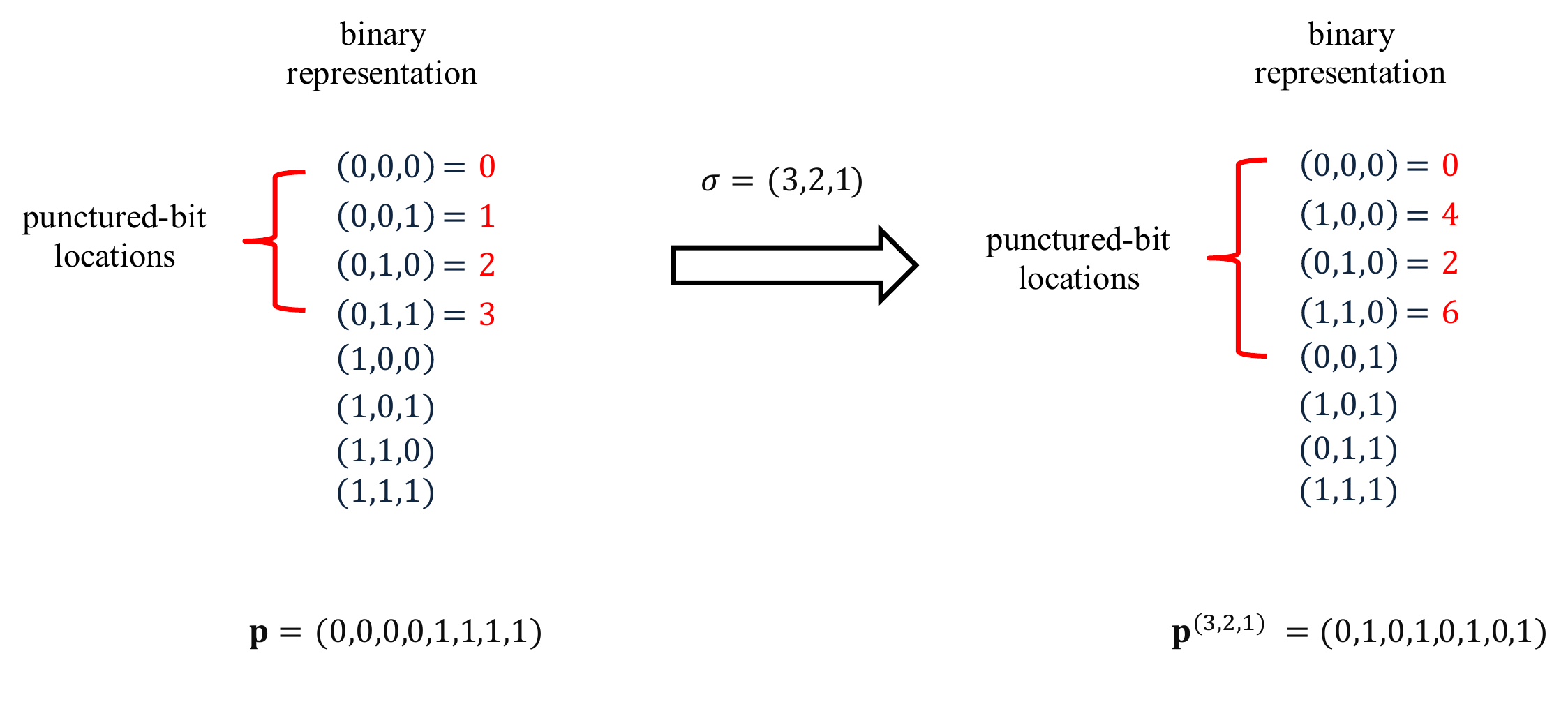}}
\caption{Construction of a {\em permuted} puncturing pattern.}
\label{fig:ppc}
\end{figure}

\subsection{Properties of reciprocal puncturing}\label{subsec:rec}

We derive the useful properties of reciprocal puncturing.  Let $\Pi_n$ denote the set of all permutations of $(1,2,...,n)$ with $|\Pi_{n}| = n!$. In the example of $n=3$, we have: 
\begin{equation*}
\Pi_{3}=\{(1,2,3),(1,3,2),(2,1,3),(2,3,1),(3,1,2),(3,2,1)\}.
\end{equation*} 

\vspace{0.2cm}
\begin{definition} For a given $\pv$ and $\sigma \in \Pi_n$, a {\em permuted} puncturing pattern $\pv^{\sigma}$ is defined with its zero-location set:
\begin{equation}
\Bc_{\pv^\sigma} = \{g^{-1}((b_{\sigma(n)}^i,...,b_{\sigma(1)}^i)): i \in \Bc_{\pv}\},\label{eq:permute}
\end{equation} where $g(i) = (b_n^{i},...,b_1^{i})$.\flushright$\blacksquare$
\end{definition} For instance, if  $\pv=(0,0,0,0,1,1,1,1)$ and $\sigma=(3,2,1)$, then  we have $\pv^{(3,2,1)} = (0,1,0,1,0,1,0,1)$ (see Fig.~\ref{fig:ppc}). With this definition, we can get:

\vspace{0.2cm}
\begin{proposition}\label{prop1} If $\pv$ is reciprocal, then $\pv^{\sigma}$ is also reciprocal for any permutation $\sigma \in \Pi_n$.
\end{proposition}
\begin{IEEEproof} The proof follows the fact that a permutation $\sigma$ in (\ref{eq:permute}) definitely preserves the both zero-inclusion and one-covering properties in Theorem~\ref{thm:rec}. 
\end{IEEEproof}

\vspace{0.2cm}
\begin{proposition}\label{prop:equi} For any reciprocal puncturing pattern $\pv$, we have:
\begin{equation}
\Gm_{\rm N}(\Bc_{\pv},\Bc_{\pv}^c) = {\bf 0}.
\end{equation}
\end{proposition}
\begin{IEEEproof} Recall that $\Gm_{\rm N}(\Bc_{\pv},\Bc_{\pv}^c)$ is the submatrix of $\Gm_{\rm N}$ by taking the columns and rows whose indices respectively belong to $\Bc_{\pv}$ and $\Bc_{\pv}^c$. Suppose there exists a non-zero element in $\Gm_{\rm N}(\Bc_{\pv},\Bc_{\pv}^c)$, i.e., $\Gm_{\rm N}(i,j) = 1$ for $i \in \Bc_{\pv}$ and $j \in \Bc_{\pv}^c$. That is, $u_i$ is added to $u_j$ to generate the $j$-th coded bit $x_j$, which shows that $i \succeq_{1} j$. Since $\pv$ is reciprocal, it should hold from Theorem~\ref{thm:rec} that any index $t$ with  $i \succeq_{1} t$ should be the element of $\Bc_{\pv}$. Accordingly, $j$ should be the element of $\Bc_{\pv}$, which is the contradiction that $j \in \Bc_{\pv}^c$. Therefore, there should be no 1's in $\Gm_{\rm N}(\Bc_{\pv},\Bc_{\pv}^c)$, i.e., $\Gm_{\rm N}(\Bc_{\pv},\Bc_{\pv}^c)={\bf 0}$. This completes the proof.
\end{IEEEproof}

\vspace{0.2cm}
\begin{remark}\label{remark:RP} From Proposition~\ref{prop:equi}, we can obtain the useful property of a reciprocal puncturing pattern $\pv$ such that assigning ``unknown" values to the synthesized channels belong to $\Bc_{\pv}$ does not impact on the performance of the punctured polar code. In the proposed RCPP code, this plays a fundamental role in exploiting an information-copy technique to generate effective information sets.\flushright$\blacksquare$
\end{remark}

In the remaining of this section, we will provide a class of reciprocal puncturing patterns which are constructed from the so-called successive puncturing pattern and a bit-wise permutation. We first define:

\vspace{0.2cm}
\begin{definition}\label{def:spec} Let $\dot{\pv}_{N_p}$ denote the {\em successive} puncturing pattern which punctures the first $N-N_p$ coded bits, namely, its zero-location set is given by
\begin{equation}
\Bc_{\dot{\pv}_{N_p}}=\{0,1,...,N-N_p-1\},
\end{equation} where $N_p$ represents the number of unpunctured coded bits. Then, we have:
\begin{itemize}
\item It is reciprocal;
\item Its permuted puncturing pattern $\dot{\pv}_{N_p}^{\sigma}$ is also reciprocal for any $\sigma \in \Pi_n$, which immediately follows from Proposition~\ref{prop1}.
\end{itemize} \flushright$\blacksquare$
\end{definition}

\vspace{0.2cm}
\begin{remark}[{QUP}] \label{remark:QUP} In particular, $\dot{\pv}_{N_p}^{(n,n-1,...,1)}$ is known as {\em quasi-uniform puncturing} (QUP) in \cite{Niu}, which can be easily verified as follows. In \cite{Niu}, the QUP algorithm proceeds as
\begin{itemize}
\item Step 1) Initialize the $\pv$ as all ones, and then, set the first $N - N_p$ bits as zeros;
\item Step 2) Perform bit-reversal permutation on the $\pv$ and then, the resulting puncturing pattern is referred to as QUP.
\end{itemize} Consider the example of $N=8$ and $N_p = 5$. Following the above steps, we have $\pv=(0,0,0,1,1,1,1,1)$ as initialization and then, by performing the bit-reversal permutation, we can get $\pv = (0,1,0,1,0,1,1,1)$. This is exactly same with  $\dot{\pv}_{5}^{(3,2,1)}$ in Definition~\ref{def:spec}. Note that, differently from this paper, the bit-reversal permutation in \cite{Niu} was applied to the encoding part. Since this only changes the indices of the input vector $\uv$, it does not changed the punctured locations. Thus,  $\dot{\pv}_{N_p}^{(n,n-1,...,1)}$  is indeed same with QUP in \cite{Niu}.\flushright$\blacksquare$
\end{remark}

\subsection{A hierarchical puncturing}\label{subsec:HC}

We define a hierarchical puncturing and derive its useful properties for the construction of a RCPP code.

\vspace{0.2cm}
\begin{definition}\label{def:HC} A reciprocal puncturing pattern $\pv$ with $w_{\rm H}(\pv)=2^{\bar{n}}=\bar{N}$ for some $\bar{n} < n$ is said to be {\em hierarchical} if 
\begin{align}
 &\uv_{\rm N}\Gm_{\rm N}(\{0,...,N-1\},\Bc_{\pv}^c)\nonumber \\
 &\;\;\;=(u_i: i \in \Bc_{\pv})\Gm_{N}(\Bc_{\pv},\Bc_{\pv}^c)+(u_i: i \in \Bc_{\pv}^c)\Gm_{N}(\Bc_{\pv}^c,\Bc_{\pv}^c)\nonumber\\
 &\;\;\;=(u_i: i \in \Bc_{\pv}^c)\Gm_{\rm \bar{N}}.\label{eq:cond}
\end{align} 
Also, since $\Gm_{N}(\Bc_{\pv},\Bc_{\pv}^c)={\bf 0}$ from Proposition~\ref{prop:equi}, the condition (\ref{eq:cond}) is equivalent to 
\begin{equation}
\Gm_{\rm N}(\Bc_{\pv}^c,\Bc_{\pv}^c) = \Gm_{\rm \bar{N}}, \label{eq:cond2}
\end{equation}where note that $|\Bc_{\pv}^c| = w_{\rm H}(\pv)=\bar{N}$.\flushright$\blacksquare$
\end{definition}

\begin{figure}
\centerline{\includegraphics[width=9cm]{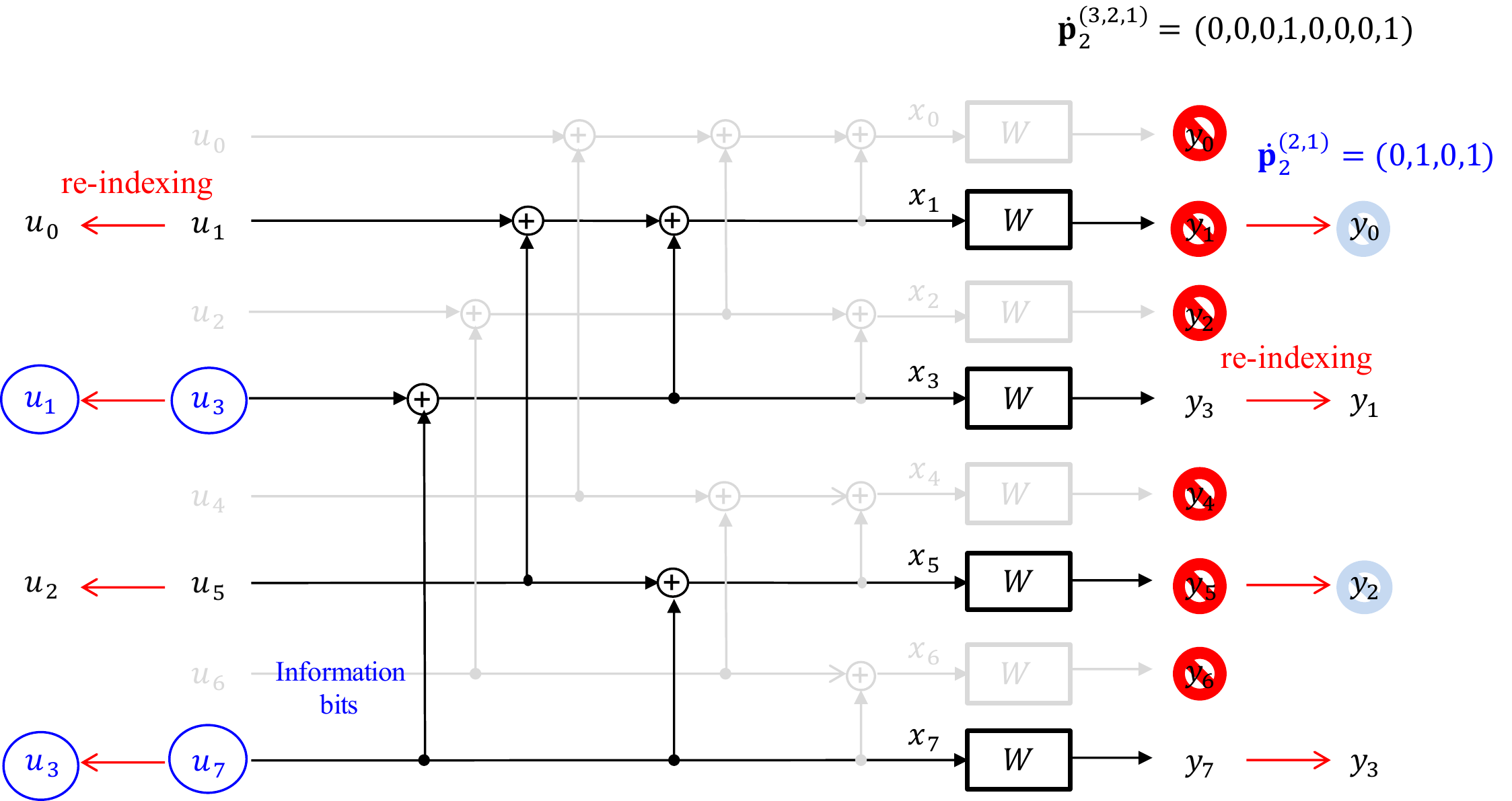}}
\caption{The punctured polar code of  $N_p = 2$, which is obtained by the puncturing pattern $\dot{\pv}^{(3,2,1)}_{2} = (0,0,0,1,0,0,0,1)$.}
\label{fig:H_PC}
\end{figure}

Starting with a simple example, we now explain two key properties of hierarchical puncturing patterns.  Consider the $N=8$, $\bar{N}=4$, and the reciprocal puncturing pattern $\dot{\pv}_{4}^{(3,2,1)}$ is used. The input-output relationship of the punctured polar code is represented as
\begin{align}
(x_i: i\in \Bc_{\dot{\pv}_{4}^{(3,2,1)}}^c)
&=\uv_{\rm N}\Gm_{\rm N}(\{0,...,7\},\Bc_{\dot{\pv}_{4}^{(3,2,1)}}^c) \nonumber\\
&=  (u_i: i \in \Bc_{\dot{\pv}_{4}^{(3,2,1)}}^c) \Gm_{\rm N} (\Bc_{\dot{\pv}_{4}^{(3,2,1)}}^c,\Bc_{\dot{\pv}_{4}^{(3,2,1)}}^c) \nonumber\\
&+ (u_i: i \in \Bc_{\dot{\pv}_{4}^{(3,2,1)}})  \Gm_{\rm N} (\Bc_{\dot{\pv}_{4}^{(3,2,1)}},\Bc_{\dot{\pv}_{4}^{(3,2,1)}}^c)\nonumber\\
  &\stackrel{(a)}{=}(u_i: i \in \Bc_{\dot{\pv}_{4}^{(3,2,1)}}^c)\Gm_{\rm \bar{N}},\label{eq:simp}
 \end{align} where (a) follows from the fact that
 \begin{align*}
 \Gm_{\rm N} (\Bc_{\dot{\pv}_{4}^{(3,2,1)}}^c,\Bc_{\dot{\pv}_{4}^{(3,2,1)}}^c) &= \Gm_{\rm \bar{N}}\\
 \Gm_{\rm N} (\Bc_{\dot{\pv}_{4}^{(3,2,1)}},\Bc_{\dot{\pv}_{4}^{(3,2,1)}}^c ) &= \bf{0}.
 \end{align*} Since it satisfies the condition  (\ref{eq:cond2}) in Definition~\ref{def:HC}, $\dot{\pv}_{4}^{(3,2,1)}$ is hierarchical. Also, we know from  (\ref{eq:A-set2}) that $(u_i: i \in \Bc_{\dot{\pv}_{4}^{(3,2,1)}}^c)$ only carry the information bits.  Let $\uv_{\rm \bar{N}} = (u_i: i \in \Bc_{\dot{\pv}_{4}^{(3,2,1)}}^c)$ and $\xv_{\rm \bar{N}}=(x_i: i \in \Bc_{\dot{\pv}_{4}^{(3,2,1)}}^c)$ (i.e., unpunctured coded bits). From (\ref{eq:simp}), we can create a length-$\bar{N}$ polar coding system as
\begin{equation}
\xv_{\rm \bar{N}}=\uv_{\rm \bar{N}} \Gm_{\rm \bar{N}},\label{eq:simp2}
\end{equation} where $\bar{N}=4 < N=8$. From the above example, we can identify the following two important properties of hierarchical puncturing patterns (which will be exploited to construct a proposed RCPP code in Section~\ref{sec:RC}):


\begin{itemize}
\item {\bf Property 1}: Information bits can be decoded using the length-$\bar{N}$ polar decoder with the (unpunctured) observation $\yv_{\rm \bar{N}}$ (see Fig.~\ref{fig:H_PC}), instead of using the original polar decoder;
\item {\bf Property 2}: Since it is also reciprocal, assigning unknown values to  the synthesized channels corresponding to $\Bc_{\dot{\pv}_{4}^{(3,2,1)}}$ does not impact on the performance of the length-$\bar{N}$ punctured polar code (see Remark~\ref{remark:RP}).
\end{itemize} 


From now on, we will provide a class of hierarchical puncturing patterns. To explain it clearly, we start with the simple case of $N=8$ and $\bar{N}=4$, i.e., half of the coded bits are punctured. As shown in Fig.~\ref{fig:PC}, the polar encoding consists of $\log{N} = 3$ levels. Also, we can easily verify that $\dot{\pv}_{4}=(0,0,0,0,1,1,1,1)$ (called successive puncturing) is hierarchical as, in this case, the following holds:
\begin{equation}
\Gm_{8}(\Bc_{\dot{\pv}_{4}}^c,\Bc_{\dot{\pv}_{4}}^c) = \Gm_{4}.
\end{equation} Also, Fig.~\ref{fig:R_PC} shows that its permuted puncturing pattern $\dot{\pv}_{4}^{(3,2,1)}=(0,1,0,1,0,1,0,1)$ is also hierarchical. From these hierarchical puncturing patterns, we identify that
\begin{equation}
b_{3}^i=0 \mbox{ for all } i \in \Bc_{\dot{\pv}_{4}} \mbox{ and } b_{1}^i = 0  \mbox{ for all } i \in \Bc_{\dot{\pv}_{4}^{(3,2,1)}},
\end{equation} where $g(i) = (b_3^i,b_2^i,b_1^i)$. In detail, the 3rd and 1st levels in  Fig.~\ref{fig:PC} are completely eliminated, respectively. Also, the remaining parts in the both cases generate the length-4 polar encoding structure, i.e., they satisfy the condition 
(\ref{eq:cond2}). Generalizing the above arguments, we can obtain that,  when half of the coded bits are punctured (e.q., $\bar{N}  =N/2$), if $b_{j}^i = 0$ for all $i \in \Bc_{\dot{\pv}_{N/2}^{\sigma}}$, then the $j$-th level in the polar encoding structure is completely eliminated, thus satisfying the  
condition (\ref{eq:cond2}). Based on this analysis, we identify that $\dot{\pv}_{N/2}^{\sigma}$ is hierarchical for any  $\sigma \in \Pi_n$, which is generalized  for any $\bar{N}=2^{\bar{n}}$ in Theorem~\ref{thm:Gen} below.

\vspace{0.2cm}
\begin{theorem}\label{thm:Gen} For any $\bar{N}=2^{\bar{n}}$ with $1 \leq \bar{n} < n$,  a puncturing pattern $\dot{\pv}_{\bar{N}}^{\sigma}$ in Definition~\ref{def:spec} is hierarchical for any $\sigma \in \Pi_{n}$.
\end{theorem}
\begin{IEEEproof}  As explained in the above, the statement holds for $\bar{n} = n-1$. Next, focusing on $\bar{n} = n - 2$, we consider the  puncturing pattern $\dot{\pv}^{\sigma}_{N/4}$ for some  $\sigma \in \Pi_{n}$. In order to use the result of $\bar{n}=n-1$, we first decompose the 
$\Bc_{\dot{\pv}^{\sigma}_{N/4}}$ into two disjoint subsets:
\begin{equation}
\Bc_{\dot{\pv}^{\sigma}_{N/4}} = \Bc_{\dot{\pv}^{\sigma}_{N/2}} \cup (\Bc_{\dot{\pv}^{\sigma}_{N/4}}\setminus \Bc_{\dot{\pv}^{\sigma}_{N/2}}).
\end{equation} Given $\Bc_{\dot{\pv}^{\sigma}_{N/2}}$, we can create the length-$N/2$ polarization structure with 
input vector $(u_i: i \in \Bc_{\dot{\pv}^{\sigma}_{N/2}}^c)$ and output vector $(x_i: i \in \Bc_{\dot{\pv}^{\sigma}_{N/2}}^c)$ (see Fig.~\ref{fig:R_PC}) since $\dot{\pv}^{\sigma}_{N/2}$ is hierarchical. Letting $N' = N/2$ and by re-indexing the input and output vectors, we can yield the length-$N'$ polar code with puncturing pattern $\dot{\pv}^{\sigma'}_{N'/2}$ where $\sigma'_{i} = \sigma_{i+1}$ for $i=1,...,n-1$  (see Fig.~\ref{fig:H_PC}). Then, $\dot{\pv}^{\sigma'}_{N'/2}$ is hierarchical with respect to the resulting length-$N'$ polar code. By combining the two stages, it is clear that $\dot{\pv}^{\sigma}_{N/4}$ is hierarchical with respect to the original length-$N$ polar code. In general for $\bar{n} = n - \ell$ with an arbitrary $1\leq \ell \leq n-1$, we can prove the statement exactly following the above procedures with $\ell$ stages. This completes the proof.
\end{IEEEproof}

We are now ready to construct a RCPP code efficiently, in which the following puncturing patterns in Remark~\ref{RC:puncturing} will be used (see Section~\ref{sec:RC} for the detailed procedures). 

\vspace{0.2cm}
\begin{remark}[{Rate-compatible, reciprocal, and hierarchical}]\label{RC:puncturing}For the construction of a proposed RCPP code, we will employ puncturing patterns $\dot{p}_{N_p}^{\sigma}$ in Definition~\ref{def:spec}. In this remark, we describe the key properties of $\dot{p}_{N_p}^{\sigma}$ as follows:
\begin{itemize}
\item For $N_1 > N_2 > \cdots > N_m$, the puncturing patterns $\dot{p}_{N_1}^{\sigma}, \dot{p}_{N_2}^{\sigma},..., \dot{p}_{N_m}^{\sigma}$ satisfy the rate-compatible constraint;
\item They are reciprocal:
\item Due to the hierarchical property, a punctured polar code, obtained by $\dot{p}_{N_i}^{\sigma}$, can be decoded using a length-$2^{\lceil \log{N_i} \rceil}$ polar code, rather than a mother polar code.
\flushright$\blacksquare$
\end{itemize}
\end{remark}

\section{The Proposed RCPP Code}\label{sec:RC}

A RCPP code consists of a family of (punctured) polar codes for which the corresponding puncturing patterns satisfy the rate-compatible constraint in (\ref{eq:R_cond}). Also, all the codes in the family should use a {\em common} information set $\Ac$ as information bits  should be unchanged during retransmissions in IR-HARQ scheme. It is extremely difficult to find an optimal common information set. Usually, it is optimized for a target code in the family  (e.g., the mother polar code or the highest-rate code). Therefore, it cannot be good for the other codes in the family, which results in a performance loss especially when a rate-change is large. Because of this, it is quite challenging to construct a good RCPP code for IR-HARQ schemes.

We address the above problem (i.e., the limitation of using a common information set) by presenting the so-called {\em information-copy} technique based on hierarchical (or reciprocal) puncturing. The main idea of the proposed RCPP code can be outlined as follows. Some information bits are repeated to frozen-bit channels as well as information-bit channels, thus yielding an {\em information-dependent} frozen vector. Here, the locations of such information bits and frozen-bit channels are determined according to rate-compatible puncturing patterns and the corresponding optimized information sets. Note that, in the encoding part, the only difference from conventional RCPP codes is that the proposed one employs the information-dependent frozen vector. In the decoding part,  {\em effective} information sets, which are optimized information sets for other codes in the family, are generated by exploiting the common information set and the information-dependent frozen vector. In this way, each code in the family can be decoded using its own optimized information set. One can concern that the use of information-dependent (unknown) frozen bits can result in a performance loss. This problem, in the proposed method, is completely avoided due to the Property 2 in  in Section~\ref{subsec:HC} of hierarchical (or reciprocal) puncturing. We provide the detailed procedures to construct the proposed RCPP code in the below.

Suppose we construct a RCPP code to send $k$ information bits with  various rates
\begin{equation}\label{eq:support_rates}
r_1=\frac{k}{N_1} < r_2=\frac{k}{N_2}< \cdots < r_m=\frac{k}{N_m}.
\end{equation} The construction method of a proposed RCPP code can be outlined as follows.

\begin{itemize}

\item {\bf step 1)} We choose the ``mother" polar code to support the above $m$ code rates as the polar code of the length  $\bar{N}_1=2^{\bar{n}_1}$, where $\bar{n}_1 = \lceil \log N_1 \rceil$.

\item {\bf step 2)} We determine a family of rate-compatible puncturing patterns,  which are denoted by the length-$\bar{N}_1$ binary vectors as
\begin{equation}
 \pv^{(1)}, \pv^{(2)}, \pv^{(3)},..., \mbox{and } \pv^{(m)},\label{eq:punc1}
 \end{equation} such that $w_{\rm H}(\pv^{(i)}) =N_i$ for $i\in \{1,...,m\}$. Each puncturing pattern $\pv^{(i)}$ generates the punctured polar code of rate $r_{i}$. Due to the rate-compatible constraint, they should satisfy
\begin{equation}
\mbox{{\bf RC-Condition:} }\Bc_{\pv^{(1)}} \subset \Bc_{\pv^{(2)}} \subset \Bc_{\pv^{(3)}} \subset \cdots \subset \Bc_{\pv^{(m)}}.\label{eq:R_cond}
\end{equation} In the proposed RCPP code,  the reciprocal (or hierarchical) puncturing patterns in Remark~\ref{RC:puncturing} are chosen as
\begin{equation}
\pv^{(i)} \eqdef \dot{\pv}^{\sigma}_{N_i}, \label{eq:rp}
\end{equation} for a fixed $\sigma \in \Pi_{\bar{n}_1}$ and for $i \in \{1,...,m\}$.

\item {\bf step 3)} We optimize a common information set $\Ac$ by taking into account the puncturing pattern $\pv^{(m)}$ (see Remark~\ref{remark:information-set} for details). Since $\pv^{(m)}$ is reciprocal, $\Ac$ should satisfy $\Ac \cap \Bc_{\pv^{(m)}} = \phi$.

\item {\bf step 4)} For simplicity,  the $m$ (punctured) polar codes in the family are denoted by $\{\Cc^{(i)}: i=1,...,m\}$, where each $\Cc^{(i)}$ is the punctured polar code, defined by
\begin{equation}
\Cc^{(i)}\eqdef\Cc(\Gm_{\bar{N}_1}, \uv_{\Ac^c}, \Ac, \pv^{(i)}= \dot{\pv}^{\sigma}_{N_i}),
\end{equation} where an {\em information-dependent} frozen vector $\uv_{\Ac^c}$ will be defined in Section~\ref{subsec:IDFV}. As explained before, this special (information-dependent) frozen vector will play a fundamental role in producing effective information sets which are good for the other codes in the family.
\end{itemize}

\vspace{0.2cm}
\begin{remark} In a work developed independently and in parallel to ours, a similar idea was introduced by Zhao et al. \cite{Zhao}. Here, the information-copy technique was used based on the extension framework. In this framework, this technique was originally proposed in earlier works in \cite{Hong-IT} and \cite{IF}, which achieves the optimal performance for a sufficiently large length (i.e., capacity-achieving). In \cite{Zhao}, the main contribution is to improve the performances of the previous works in \cite{Hong-IT} and \cite{IF} for practical finite lengths, by entangling the subcodes (corresponding generator matrices) appropriately. Whereas, in this work, we apply the information-copy technique to puncturing framework in a non-trivial way. Also, it is remarkable that our work based on the successive puncturing is equivalent to that in  \cite{Zhao}, by simply changing the construction order. Therefore, we can say that this work extends the independent work in \cite{Zhao}, by generating more puncturing patterns suitable for the information-copy technique. \flushright$\blacksquare$
\end{remark}

In the following subsections, we will explain how to construct an information-dependent frozen vector and the encoding/decoding procedures of the proposed RCPP code. We first provide the notations for information sets which will be employed in the sequel.

\vspace{0.2cm}
\begin{definition} We let $\Tc^{(i)}$ denote the optimal information set for each code $\Cc^{(i)}$ in the family. Note that this information set is independently optimized by taking into account an associated puncturing pattern (see Remark~\ref{remark:information-set}). Due to a certain requirement (will be explained below), the effective information set (obtained by the information-copy technique) can be different from the optimal one. To clarify such difference, the effective information set is denoted by $\Ac^{(i)}$. Note that the code $\Cc^{(i)}$ in the family is decoded with the effective information set $\Ac^{(i)}$. These notations will be exploited in the below.
\flushright$\blacksquare$
\end{definition}

\subsection{Information-dependent frozen vector}\label{subsec:IDFV}

\begin{figure*}
\centerline{\includegraphics[width=17cm]{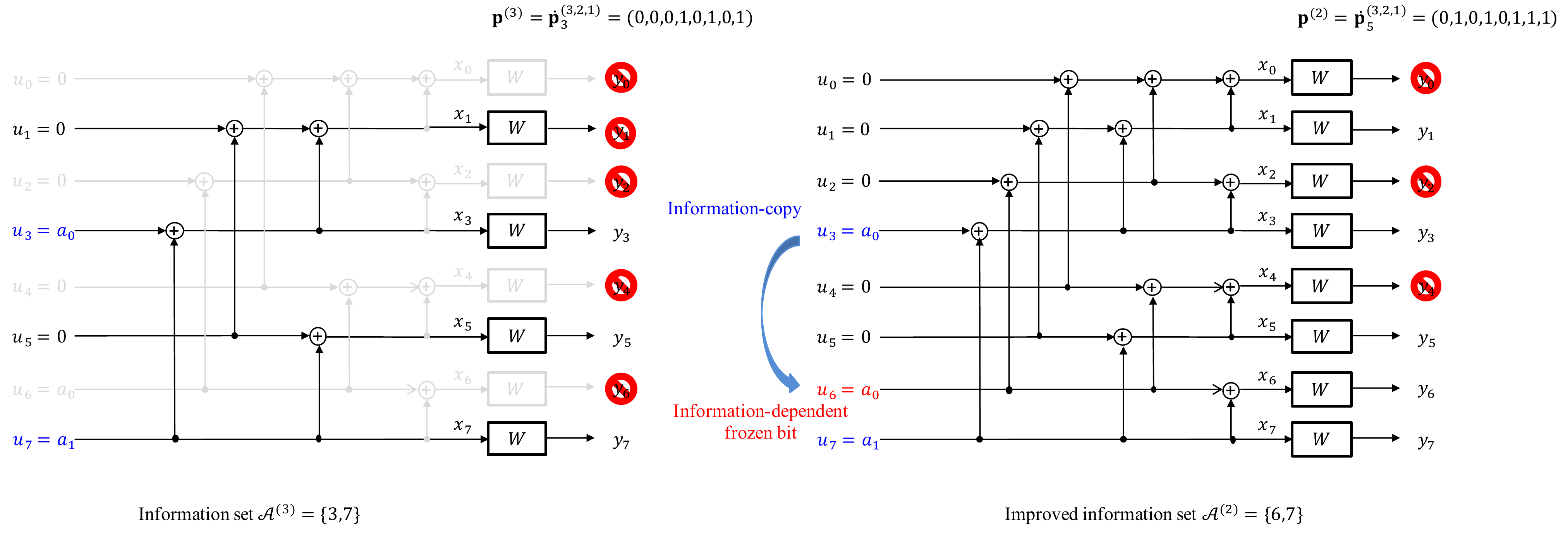}}
\caption{Illustration of a proposed information-copy technique. Note that the decoding order of SC decoder is given by $u_0 \rightarrow u_4 \rightarrow u_2 \rightarrow u_6 \rightarrow u_1 \rightarrow u_5 \rightarrow u_3 \rightarrow u_7$.}
\label{fig:REP}
\end{figure*}

In this section, we present an information-copy technique which produces an information-dependent frozen vector (equivalently, {\em effective} information sets $\Ac^{(i)}$ for the codes  $\Cc^{(i)}$, $i=1,...,m-1$). We will explain the main idea with the simple case of $k=2$ and
\begin{equation}
r_1 = \frac{2}{8} < r_2 = \frac{2}{5} < r_3=\frac{2}{3}.
\end{equation} From (\ref{eq:rp}) in Step 2), the rate-compatible puncturing patterns are chosen as
\begin{align}
\pv^{(1)}&=\onev=(1,1,1,1,1,1,1,1)\\
\pv^{(2)}&= \dot{\pv}^{(3,2,1)}_{5}=(0,1,0,1,0,1,1,1)\\
\pv^{(3)}&= \dot{\pv}^{(3,2,1)}_{3}=(0,0,0,1,0,1,0,1).
\end{align} In this example, the corresponding optimal information sets are obtained as
\begin{equation}
\Tc^{(1)}=\{6,7\}, \Tc^{(2)}=\{6,7\}, \mbox{ and } \Tc^{(3)}=\{3,7\}.
\end{equation}
As in Step 3), the common information set $\Ac$ is given by
\begin{equation}
\Ac=\Ac^{(3)}=\Tc^{(3)}=\{3,7\}.
\end{equation}  We can see that $\Ac=\Tc^{(3)}$ is not optimal information set for the code $\Cc^{(2)}$ because $\Tc^{(2)}=\{6,7\} \neq \Tc^{(3)}$, i.e., $I(W^{(6)}_{\pv^{(2)}}) > I(W^{(5)}_{\pv^{(2)}})$ (see Fig.~\ref{fig:REP}). Nevertheless, in conventional RCPP codes, all the codes in the family use $\Ac$ as the information sets. In the proposed RCPP code, however, each code in the family is able to use its own optimized information set. Namely, 
$\Cc^{(2)}$ can use $\Tc^{(2)}=\{6,7\}$ as its information set. In the below, we will show how it works.

Fig.~\ref{fig:H_PC} shows that  $\Cc^{(3)}$ can be decoded via the length-4 polar code $\Cc(\Gm_4, \{1,3\},(0,0), \pv=(0,1,1,1))$, instead of  the mother polar code $\Cc(\Gm_{8},\{3,7\},\uv_{\Ac^c},\onev)$. This is possible  because $\dot{\pv}_{3}^{(3,2,1)}$ is hierarchical (see Property 1 in Section~\ref{subsec:HC}). It is also remarkable that  the frozen-bit channels belong to $\Bc_{\pv^{(3)}}=\{0,2,4,6\}$ are not associated with this decoding, which makes it possible to allocate ``unknown" frozen bits to those frozen-bit channels without degrading the performance of  $\Cc^{(3)}$ (see Property 2 in Section~\ref{subsec:HC}). Hence, the information-bit  $u_3=a_0$ can be copied to the frozen-bit channel 6 (i.e., $u_6 = u_3=a_0$). In the decoding of $\Cc^{(2)}$, $u_5$ can operate as a frozen bit since the copied one $u_6$ is decoded earlier. This shows that $\Cc^{(2)}$ can be decoded using the {\em effective} information set $\Ac^{(2)}=\Tc^{(2)}=\{6,7\}$, instead of $\Ac=\{3,7\}$. Next focus on the decoding of $\Cc^{(1)}$. As before, we copy the information bit $a_0$ to the frozen-bit channel 4 (i.e., $u_6 = u_3 = u_4 =a_0$). In the decoding of $\Cc^{(1)}$, both $u_3$ and $u_6$ can operate as frozen bits since the copied one $u_4$ is decoded earlier. Also, from  Property 2 in Section~\ref{subsec:HC}, the copied bit $u_6$ does not impact on the decoding of both $\Cc^{(2)}$ and $\Cc^{(3)}$. Thus, $\Cc^{(1)}$ can be decoding using the {\em effective} information set $\Ac^{(1)}=\Tc^{(1)}=\{6,7\}$. In this simple example, we know that $\Ac^{(i)} = \Tc^{(i)}$ but it is not necessary.  Consequently, the information-dependent frozen vector is obtained as
 \begin{equation}
\uv_{\Ac^{c}}=(u_0,u_1,u_2,u_4,u_5,u_6) = (0,0,0,u_3=a_0,0,u_3=a_0). 
\end{equation}
Accordingly, the punctured polar codes in the family are defined as $\Cc^{(i)} = \Cc(\Gm_{8},\Ac=\{3,7\},\uv_{\Ac^c}=(0,0,0,u_3=a_0,0,u_3=a_0), \pv^{(i)})$
for $i=1,2,3$. From this, we can see that an input vector $\uv$ (i.e., $\uv_{\Ac}$ and $\uv_{\Ac^c}$) of the mother polar code is determined as
\begin{equation}\label{eq:precoding}
(a_0,a_1) \underbrace{\left[ \begin{array}{cccccccc}
      0 &  0 & 0 & 1 & 1 & 0 & 1 & 0\\
      0 &  0 & 0 & 0 & 0 & 0 & 0 & 1
  \end{array}  \right]}_{\eqdef \Pm}=(u_0,u_1,...,u_{\bar{N}_1-1}),
\end{equation} where $(a_0,a_1)$ represents the two information bits and the $|\Ac|\times \bar{N}_1$ matrix $\Pm$ is called precoding matrix. Note that $\Pm$ is constructed as a function of  $\uv_{\Ac^c}$  and $\Ac$, where $\Pm(0,3)=\Pm(1,7)=1$ for $\Ac^{(3)} = \{3,7\}$, $\Pm(0,6)=1$ for $\Ac^{(2)} \setminus \Ac^{(3)}= \{6\}$, and $\Pm(0,4)=1$ for $\Ac^{(1)}\setminus \Ac^{(2)}=\{4\}$.  Furthermore, we identify that, in order not to affect the performances of the other codes in the family, information-copy technique should be performed only when the following requirement holds:
\begin{itemize}
\item {\bf IC-Requirement:} The indices of frozen-bit channels to be copied, in order to construct $\Ac^{(i)}$ from a given $\Ac^{(i+1)}$, should belong to $(\Tc^{(i)}\setminus \Ac^{(i+1)}) \cap \Bc_{\dot{\pv}_{N_{i+1}}^{(3,2,1)}}$, for $i=1,2$. Note that the intersection with $\Bc_{\dot{\pv}_{N_{i+1}}^{(3,2,1)}}$ is required to ensure that the copied bits do not affect the performance of $\Cc^{(i+1)}$ (see Property 2 in Section~\ref{subsec:HC}).
\end{itemize}

\vspace{0.2cm}
\begin{example} In this example, it is shown that, due to IC-requirement, some information bits cannot be copied and accordingly, $\Tc^{(i)} \neq \Ac^{(i)}$. Suppose that 
\begin{align*}
\pv^{(1)} &= \dot{\pv}^{(3,2,1)}_{8}=(1,1,1,1,1,1,1,1)\\
\pv^{(2)} &= \dot{\pv}^{(3,2,1)}_{5}=(0,1,0,1,0,1,1,1),
\end{align*} and the associated optimal information sets are given by $\Tc^{(1)} = \{4,5\}$ and $\Tc^{(2)} = \Ac^{(2)} = \{3,7\}$, respectively. From $\pv^{(2)}$, we have that $\Bc_{\dot{\pv}_{5}^{(3,2,1)}}=\{0,2,4\}$. From IC-Requirement,   $\Ac^{(2)}$ cannot include the index $5$ as $5 \notin \Bc_{\dot{\pv}_{5}^{(3,2,1)}}$. Hence, we obtain that  $\Ac^{(1)} = \{ 4,7\}  \neq \Tc^{(2)}$. \flushright$\blacksquare$
\end{example}

Taking the above IC-requirement into account, we provide the systematic algorithms to produce {\em effective} information sets and a precoding matrix $\Pm$ (see Algorithms 1 and 2, respectively). It is noticeable that the precoding matrix $\Pm$ is used at the encoding side and the effective information sets are used at the decoding side. We briefly explain the main idea of these algorithms as follows. Consider the IR-HARQ scheme to support the $m$ code rates in (\ref{eq:support_rates}), i.e., $r_{1}=\frac{k}{N_1}<\cdots < r_m = \frac{k}{N_m}$. Then, we have:
\begin{itemize}
\item As explained in Remark 1, the $m$ information sets $\Tc^{(1)}, \Tc^{(2)},...,$ and $\Tc^{(m)}$ are optimized by  taking the corresponding puncturing patterns into account (i.e., based on $I(W^{(j)}_{\pv^{(i)}})$). 
\item From  $\Tc^{(i)}$'s, Algorithm 1 produces effective information sets $\Ac^{(i)}$ for $i=1,...,m$. Note that $\Ac^{(i)}$ is not always identical to $\Tc^{(i)}$.
\item Also, from $\Ac^{(i)}$'s, Algorithm 2 generates a precoding matrix $\Pm$ (i.e., information-dependent frozen vector).
\end{itemize}

\begin{algorithm}[h]\label{alg:eff_A}
\caption{Effective Information Sets $\Ac^{(i)}$'s}
 {\bf Input:} Optimized information sets $\Tc^{(i)}$ for $i\in \{1,...,m\}$ and set $\Ac =\Ac^{(m)}= \Tc^{(m)}$.

 {\bf Output:} Optimized (effective) information sets $\Ac^{(j)}$ for $j\in\{1,...,m\}$.
 \vspace{0.1cm}
  
 {\bf Algorithm:}
   
\;\;\; For $j=m-1,...,1$

 \begin{enumerate}
 \item Let  
 \begin{align*}
 \Ic_1 &= (\Tc^{(j)} \setminus \Ac^{(j+1)}) \cap \Bc_{\dot{\pv}^{\sigma}_{N_{j+1}}}\eqdef\{\ell_1,...,\ell_{|\Ic_1|}\}\\
 \Ic_2 &= \mbox{min-ind}^{(|\Ic_1|)}\{I(W^{(i)}_{\pv^{(j)}}): i \in \Ac^{(j+1)}\},
 \end{align*} where $\psi(\ell_1) <  \cdots < \psi(\ell_{|\Ic|})$ with the bit-reverse permutation $\psi(\cdot)$.
 \item  Information-copy set $\Ic_{c}$:
 \begin{itemize}
 \item Initialization: $\Ic_{c} = \phi$ and $\Ic_{d} = \phi$. 
 \item For $i=1,...,|\Ic_1|$
 \begin{enumerate}
 \item $\Sc = \{q \in \Ic_{2}\setminus\Ic_{d} : \psi(q) >  \psi(\ell_{i})\}$
 \item If $\Sc\neq \phi$, then
 \begin{equation*}
 \Ic_{c} =\Ic_{c}\cup\{\ell_{i}\} \mbox{ and } \Ic_{d} =\Ic_{d}\cup\{q^*\},
 \end{equation*} where $q^* =\min_{q \in \Tc} \psi(q)$. 
 \end{enumerate}
 \end{itemize}
 \item Effective information set:  $\Ac^{(j)} \eqdef \Ic_{c} \cup (\Ac^{(j+1)}\setminus \Ic_{d})$. 
 \end{enumerate}
 \end{algorithm}

\begin{algorithm}[h]\label{alg:input}
\caption{Precoding Matrix $\Pm$}
 
 {\bf Input:}  The effective information sets $\Ac^{(j)}$ for $j \in \{1,...,m\}$ and length $\bar{N}_1$.

 {\bf Output:}  The $|\Ac| \times \bar{N}_1$ precoding matrix $\Pm$ where $\Pm(i,j)$ denotes the $(i,j)$-th element of  $\Pm$ for $i=0,1,...,|\Ac|-1$ and $j=0,1,...,\bar{N}_1 -1$.
\vspace{0.1cm}

 {\bf Initialization:} 
 \begin{itemize}
 \item $\Ac^{(m)} = \Ac = \left\{\ell_{0},....,\ell_{|\Ac^{(m)}|-1}\right\}$ and $\Pm={\bf 0}$.
 \item Define a mapping $h: \Ac^{(m)} \rightarrow \{0,1,...,|\Ac^{(m)}|-1\}$, i.e., $h(\ell_j) = j \mbox{ for } j=0,...,|\Ac^{(m)}|-1$.
 \item Let $\Ac^{(j)} \setminus \Ac^{(j+1)} \eqdef \{\ell^{(j)}_{1},...,\ell^{(j)}_{d_j}\}$ for $j\in \{1,...,J-1\}$.
 \end{itemize}
 
 {\bf Algorithm:}
 \begin{itemize}
\item Assign $\Pm(h(\ell_j),\ell_j) = 1$ for $j=0,...,|\Ac^{(m)}|-1$.

\item For $j=m-1,...,1$

$\;\;\;\;\;$ Assign  $\Pm(h( i_{t}^{(j)}), i_{t}^{(j)})= 1$ for $t \in [1:d_j]$.

 \end{itemize}
\end{algorithm}

\begin{figure*}
\centerline{\includegraphics[width=15cm]{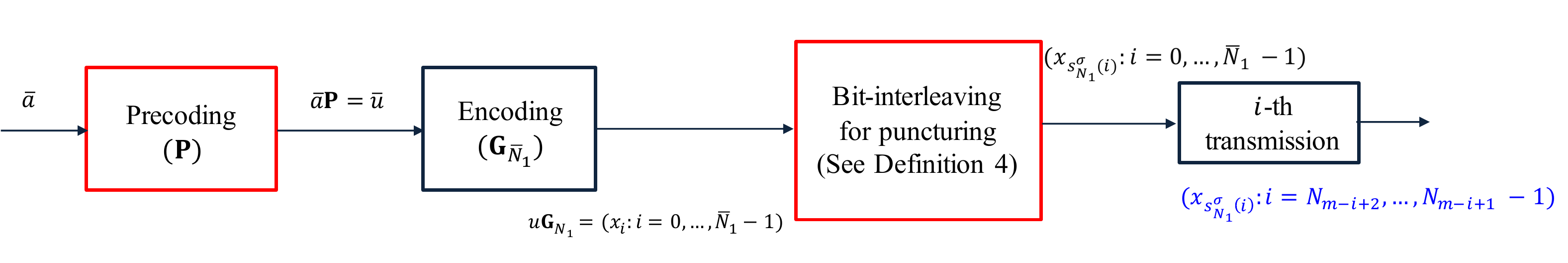}}
\caption{Encoding structure of the proposed RCPP code in which the bit-interleaving is used to produce a rate-compatible puncturing pattern (see Definition 4 for details).}
\label{fig:enc}
\end{figure*}


\subsection{Encoding and decoding}

We describe the encoding and decoding procedures of the proposed RCPP code. First of all, we simplify the expression of the proposed rate-compatible puncturing patterns in (\ref{eq:rp}), by introducing the so-called {\em seed sequence}.

\vspace{0.2cm}
\begin{definition}\label{def:seed} Given $\bar{N}_1$ and $\sigma$, we define a seed sequence by the following length-$\bar{N}_1$ binary vector
\begin{equation}
\sv_{\bar{N}_1}^{\sigma}\eqdef (g^{-1}(b^0_{\sigma(3)},b^0_{\sigma(2)},b^0_{\sigma(1)}),...,g^{-1}(b^{\bar{N}_1-1}_{\sigma(3)},b^{\bar{N}_1-1}_{\sigma(2)},b^{\bar{N}_1-1}_{\sigma(1)})),\label{eq:seed}
\end{equation}where $(b^{i}_n,...,b^{i}_1)$ denotes the binary representation of `$i$' for $i=0,...,\bar{N}_1-1$. Then, the puncturing pattern $\pv^{(i)}= \dot{\pv}^{\sigma}_{N_i}$ in (\ref{eq:rp}) is simply defined with its zero-location set as
\begin{equation}
\Bc_{\pv^{(i)}} = \{\sv_{\bar{N}_1}^{\sigma}(i): i = 0,1,...,N_i-1\}.\label{eq:seed_set}
\end{equation}\flushright$\blacksquare$
\end{definition}

\vspace{0.2cm}
\begin{example} Suppose $\bar{N}_1  = 8$ and $\sigma=(3,2,1)$ (e.g., QUP). From (\ref{eq:seed}), we can easily obtain the length-$\bar{N}_1$ seed sequence as
\begin{equation}
\sv_{8}^{(3,2,1)}=(0,4,2,6,1,5,3,7).
\end{equation} For $N_1 = 7$, $N_2=5$, and $N_3 =3$, the corresponding rate-compatible puncturing patterns are obtained from (\ref{eq:seed_set}) as
$\Bc_{\pv^{(1)}} = \{0\}$, $\Bc_{\pv^{(2)}} = \{0,2,4\}$, and $\Bc_{\pv^{(3)}} = \{0,1,2,4,6\}$. Accordingly, given an encoded output $(x_{i}: i=0,...,7)$, IR-HARQ scheme proceeds as follows:
\begin{itemize}
\item 1st transmission:  $(x_{\sv_{8}^{(3,2,1)}(i)}: i=0,..,N_{3}-1)=(x_{3},x_{5},x_{7})$;
\item 2nd transmission: $(x_{\sv_{8}^{(3,2,1)}(i)}: i=N_{3},..,N_{2}-1)=(x_{1},x_{6})$;
\item 3rd transmission: $(x_{\sv_{8}^{(3,2,1)}(i)}: i=N_{2},..,N_{1}-1)=(x_{2},x_{4})$.
\end{itemize}\flushright$\blacksquare$
\end{example} Generalizing the above example, we have:

\vspace{0.2cm}
{\bf Encoding:} Consider the proposed RCPP code to support $m$ code rates in (\ref{eq:support_rates}). We first generate an input vector $\uv$ using a precoding matrix $\Pm$ and then, produce a polar-encoded output $\xv$. Then, at the $i$-th (re)transmission, the following part of the encoded output $\xv=\uv\Gm_{\bar{N}_1}$ is transmitted:
\begin{equation}
(x_{\sv_{\bar{N}_1}^{\sigma}(i)}: i=N_{m-i+2},...,N_{m-i+1}-1),
\end{equation} for $i=1,...,m$ and with initial value $N_{m+2}= 0$. Then, the overall encoding structure is illustrated in Fig.~\ref{fig:enc}.

\vspace{0.2cm}
\begin{remark} As shown in Fig.~\ref{fig:enc}, the proposed RCPP code seems to have an additional precoding operation, compared with conventional RCPP codes. However, we would like to emphasize that this precoding is also performed (implicitly) in conventional RCPP codes (i.e., mapping information bits to an input vector). In this case, each row of $\Pm$ has only one 1's due to the use of all-zero frozen vector while in the proposed RCPP code, some rows of $\Pm$ can have more than one 1's because of information-copy technique. Clearly, this minor difference does not affect the overall encoding complexity. \flushright$\blacksquare$
\end{remark}

\vspace{0.2cm}
{\bf Decoding:} Consider the decoding of the proposed RCPP code after the $i$-th (re)transmission. In conventional RCPP codes, the polar decoding is performed via the mother polar code $\Cc^{(1)}=\Cc(\Gm_{\bar{N}_1}, \uv_{\Ac^c}, \Ac, \pv^{(1)})$. Whereas, as shown in 
Fig.~\ref{fig:REP}, the proposed RCPP code can be decoded with the length-$\bar{N_i}$  (possibly shorter) polar code, defined by
\begin{equation}
\Cc(\Gm_{\bar{N}_i}, \uv_{\bar{\Ac}^{c}},\bar{\Ac},\pv), 
\end{equation} where $\bar{\Ac}$ (i.e., {\em effective} information set), $\uv_{\bar{\Ac}^{c}}$, and $\pv$ will be specified from the given $\pv^{(i)}$, $\Ac^{(i)}$, and $\uv_{\Ac^c}$. 
Recall that $\uv_{\Ac^c}$ is able to contain the ``unknown" values. For the example in (\ref{eq:precoding}), $\uv_{\Ac^c} = (u_0=0, u_1=0, u_2=0, u_4=0, u_5=0, u_6=u_3)$ has the unknown frozen bit $u_6=u_3$. In the decoding, if one of $u_3$ and $u_6$ are decoded, the other bit is immediately copied and operates as the ``known" frozen-bit. For the ease of expression, we define 
\begin{equation}\label{eq:l}
\ell^{(i)}_j \eqdef \mbox{ the $(j+1)$-st smallest index in } \Bc_{\dot{\pv}^{\sigma}_{\bar{N}_i}}^c,
\end{equation}  for $j=0,...,|\Bc_{\dot{\pv}^{\sigma}_{\bar{N}_i}}^c|-1$. As seen in Fig.~\ref{fig:H_PC}, using the {\em re-indexing}, we obtain:
\begin{align}\label{eq:bar_A}
\bar{\Ac}&=\{j: \ell_j^{(i)} \in \Ac^{(i)}\} \nonumber\\
\pv&=(p_{t}^{(i)}: t \in \Bc_{\dot{\pv}^{\sigma}_{\bar{N}_i}}^c)\nonumber\\
\uv_{\bar{\Ac}^c} &= (u_t: t \in \Bc_{\dot{\pv}^{\sigma}_{\bar{N}_i}}^c \setminus \Ac^{(i)}),
\end{align}where $\pv^{(i)}=\dot{\pv}^{\sigma}_{N_i}\eqdef(p_{0}^{(i)},...,p_{\bar{N}_1-1}^{(i)})$.

\vspace{0.2cm}
\begin{example} We revisit the simple case in Section~\ref{subsec:IDFV} and focus on the polar decoding after 3rd transmission. In this case, we have that $\Ac^{(3)}=\{3,7\}$, $\pv^{(3)} = \dot{\pv}_{3}^{(3,2,1)}(0,0,0,1,0,1,0,1)$, and $\uv_{\Ac^c} = (u_0=0, u_1=0, u_2=0, u_4=0, u_5=0, u_6=u_3)$. When $\bar{N}_3 =4$, we obtain $\Bc_{\dot{\pv}^{(3,2,1)}_{4}}^c=\{1,3,5,7\}$ and, using (\ref{eq:l}), we have
\begin{equation*}
\ell_0^{(3)} = 1, \ell_1^{(3)} = 3, \ell_2^{(3)} = 5, \mbox{ and }\ell_3^{(3)} = 7.
\end{equation*} From~(\ref{eq:bar_A}), we then get
\begin{equation*}
\bar{\Ac}=\{1,3\}, \pv=(0,1,1,1), \mbox{ and } \uv_{\bar{\Ac}^c}=(u_1=0,u_5=0).
\end{equation*} In this case, the decoding is performed via  
\begin{equation}
\Cc(\Gm_4, \uv_{\bar{\Ac}^c}=(0,0), \bar{\Ac}=\{1,3\}, \pv=(0,1,1,1)).
\end{equation} \flushright$\blacksquare$
\end{example}

\section{Simulation Results for IR-HARQ Scheme} \label{sec:SIM}

\begin{figure}
\centerline{\includegraphics[width=10cm]{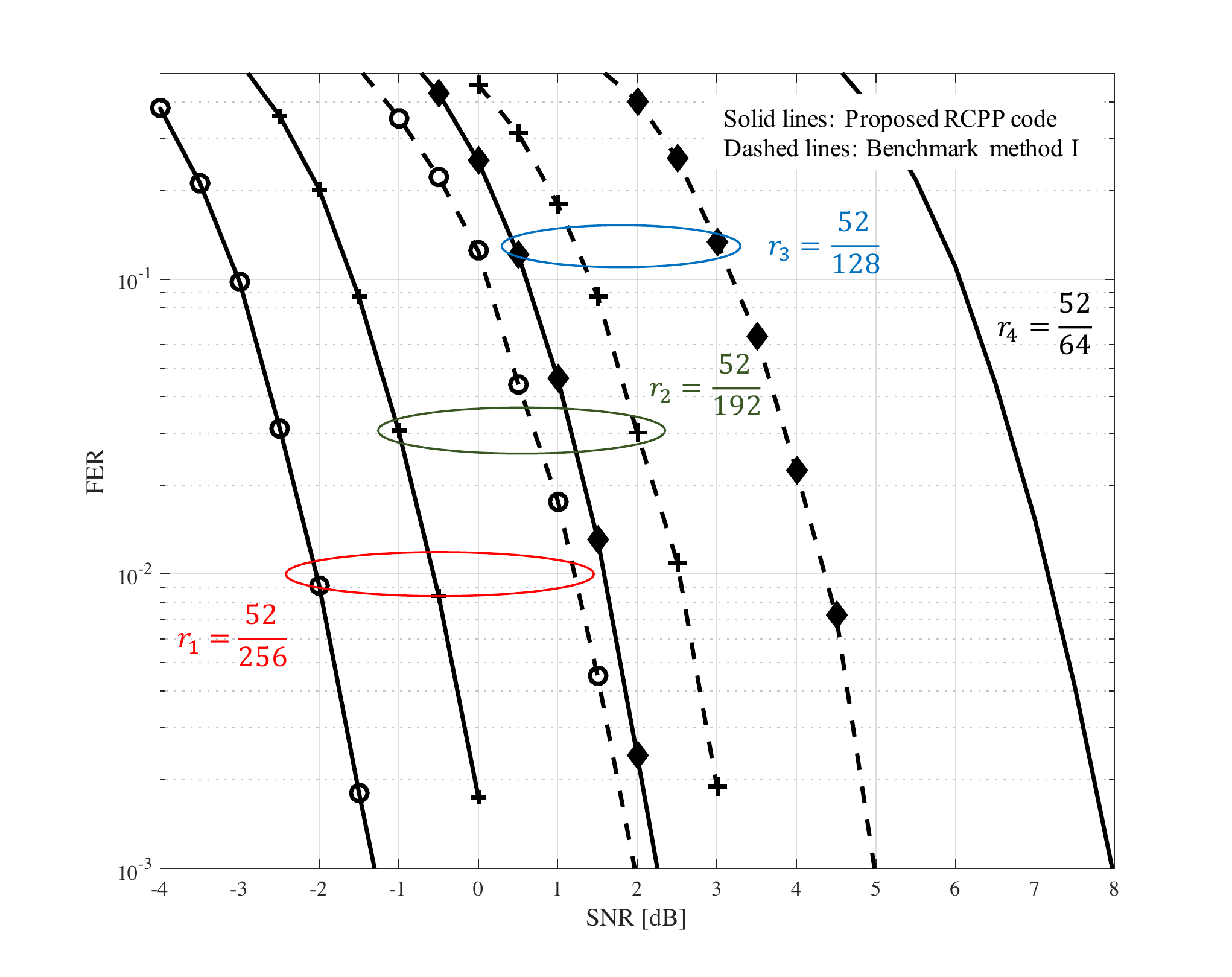}}
\caption{Performance comparison of the proposed method and benchmark method I to support the code rates $r_1=\frac{52}{256} < r_2 = \frac{52}{192} < r_3 = \frac{52}{128} < r_4 = \frac{52}{64}$. Both methods employ the identical rate-compatible puncturing patterns (known as QUP) and the common information set $\Ac$ (which is optimized for the highest-rate code). The only difference is that the proposed approach uses the information-dependent frozen vector while the benchmark scheme uses the conventional all-zero frozen vector.}
\label{fig:sim}
\end{figure}

\begin{figure}
\centerline{\includegraphics[width=10cm]{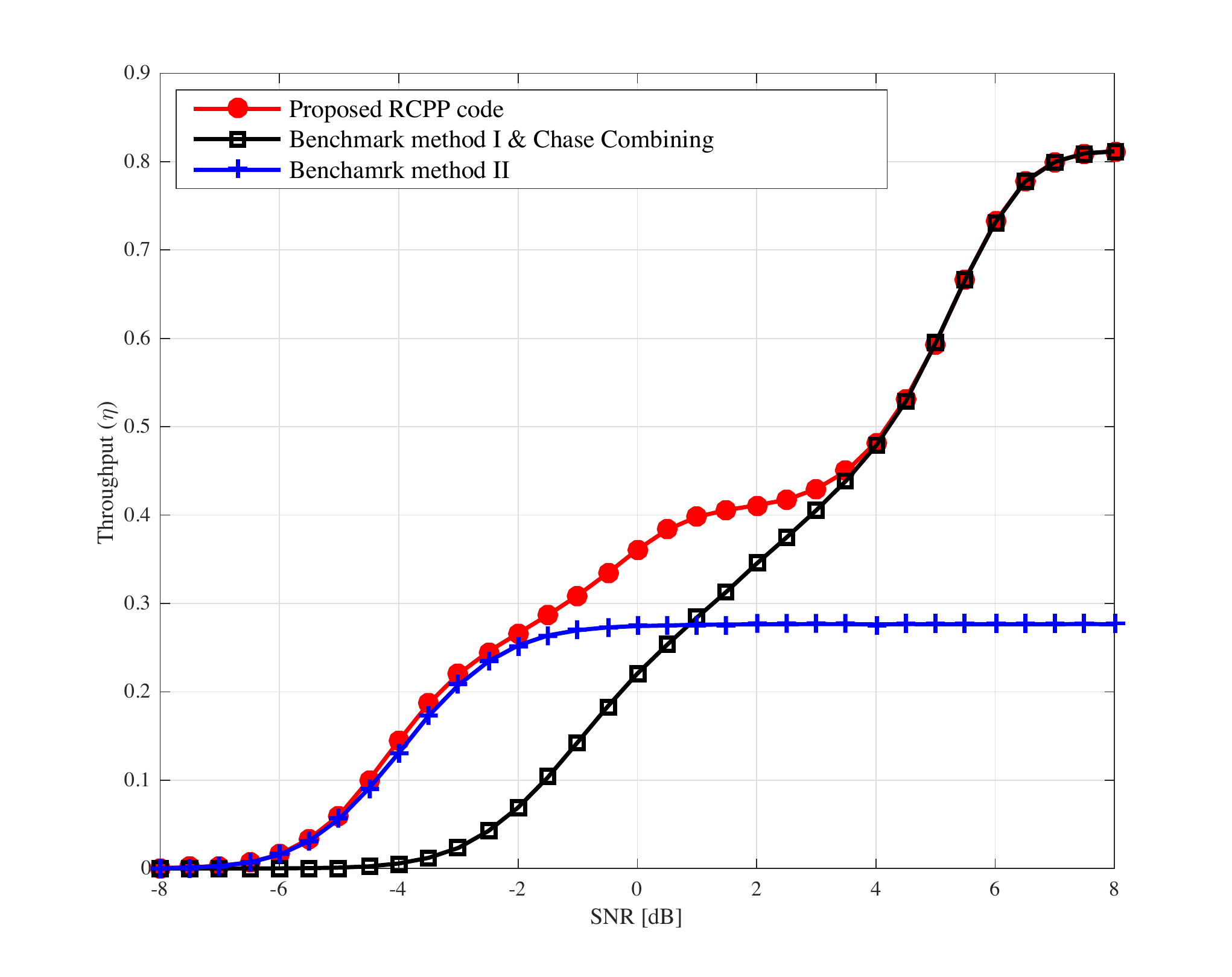}}
\caption{Throughput for various IR-HARQ schemes. All three methods support the code rates  $r_1=\frac{52}{256} < r_2 = \frac{52}{192} < r_3 = \frac{52}{128} < r_4 = \frac{52}{64}$, and employ the identical rate-compatible puncturing patterns (known as QUP). The proposed method and benchmark method I use the same common information set which is optimized for the highest-rate code while benchmark method II uses the common information set which is optimized for the lowest-rate code. As before, only the proposed method employs the information-dependent frozen vector.}
\label{fig:throughput}
\end{figure}

In this section, we consider the IR-HARQ scheme based on the proposed RCPP code to  send $k=52$ information bits via four different code rates as
\begin{equation*}
r_1=\frac{52}{256} < r_2=\frac{52}{192} < r_3=\frac{52}{128} < r_4=\frac{52}{64}.
\end{equation*}  At the decoding part, the list decoder with list-size 8 and 8-bit CRC are used. Note that, thus, the actual size of information bits, with respective to (punctured) polar codes, is equal to 60. The proposed RCPP code is constructed as follows.
\begin{itemize}
\item The  rate-compatible puncturing patterns (e.g., QUP) are chosen as
\begin{align}
&\pv^{(1)}=\onev, \pv^{(2)} = \dot{\pv}^{(8,7,...1)}_{192},  \nonumber\\
&\pv^{(3)} = \dot{\pv}^{(8,7,...,1)}_{128}, \mbox{ and } \pv^{(4)} = \dot{\pv}^{(8,7,...,1)}_{64}.\label{ex-pun}
\end{align} 
\item Using the optimization method in Remark~\ref{remark:information-set}, we separately find the optimal information set $\Tc^{(i)}$ for each code $\Cc^{(i)}$ for $i=1,2,3,4$. Here, each $\Tc^{(i)}$ is associated with the puncturing pattern $\pv^{(i)}$ for $i=1,2,3,4$. From them, we choose the common information set as $\Ac = \Ac^{(4)} = \Tc^{(4)}$.
\item Using Algorithm 1, we obtain the effective information sets $\Ac^{(i)}$ for $i=1,2,3$.  Also, using $\Ac^{(i)}$'s and Algorithm 2, we obtain the precoding matrix $\Pm$.
\item At the decoding side, the (punctured) polar code $\Cc^{(i)}$ is decoded using the {\em effective} information set 
$\Ac^{(i)}$ for $i=1,2,3,4$. From Remark~\ref{RC:puncturing}, furthermore, both $\Cc^{(1)}$ and $\Cc^{(2)}$ are decoded using the length-256 polar code, $\Cc^{(3)}$ is using the length-128 polar code, and $\Cc^{(4)}$ is using the length-64 polar code.
\end{itemize}


To show the superiority of the proposed method, we compare the performances with the benchmark methods. They employ the same rate-compatible puncturing patterns  in (\ref{ex-pun}) (e.g., QUP) while all the codes in the family are decoded only using the common information set $\Ac$, namely, the information-copy technique is not employed. Here, the common information sets are optimized in the two different ways:
\begin{itemize}
\item {\bf Benchmark method I}: The common information set is optimized for the {\em highest-rate} code in the family (i.e., $\Ac = \Tc^{(4)}$).
\item {\bf Benchmark method II}: The common information set is optimized for the {\em lowest-rate} code in the family  (i.e., $\Ac = \Tc^{(1)}$).
\end{itemize} It is noticeable that, when $\Ac=\Tc^{(i)}$, the code $\Cc^{(j)}$ for some $j > i$ can suffer from a severe error-floor since the common information set does not ensure that $\Bc_{\pv^{(j)}} \cap \Tc^{(i)} \neq \phi$ for some $j > i$. In this example, we observed that if we choose $\Ac = \Tc^{(i)}$ for some $i < 4$, the code $\Cc^{(4)}$ (in the benchmark method) suffers from a severe error-floor, which yields a lower throughput for IR-HARQ scheme  (see the benchmark method II in Fig.~\ref{fig:throughput}). From Fig.~\ref{fig:sim}, we observe that the proposed RCPP code can significantly outperform the benchmark method. As expected, the performance gain becomes lager as a code rate is lower. The corresponding throughputs for IR-HARQ schemes are provided in Fig.~\ref{fig:throughput}. Here, we also considered {\em chase combining} (CC) HARQ scheme in which the highest-rate polar code of rate $r_4$ is repeated for every retransmission. Fig.~\ref{fig:sim} shows that the proposed RCPP code achieves higher throughput than the benchmark schemes. Although in this particular example, the throughput gain of the proposed method is only attained at lower SNRs, larger gains at higher SNRs can be made by choosing a family of codes with more high-rate cods. Moreover, we observe that, without leveraging the proposed information-copy technique, IR-HARQ scheme cannot get a throughput gain over the simple CC-HARQ scheme. Thus, the proposed method would play a fundamental role in constructing a RCPP code for IR-HARQ schemes. Here, the (expected) throughput ($\eta$) is computed as
\begin{equation*}
\eta = \frac{k(1-p_f)}{\sum_{i=1}^{m-1} N_{m-(i-1)}(1-p_i)\prod_{j=1}^{i-1}p_j+N_1\prod_{j=1}^{m-1}p_j},
\end{equation*} where $p_i$ represents the probability of a frame error on the $i$-th transmission, given that all the previous transmissions failed, for $i=1,...,m-1$, and $p_f$ represents the probability that none of the transmission succeeded. Recall that $N_{m-(i-1)}$ denotes the number of transmitted coded bits until the $i$-th transmission (see (\ref{eq:support_rates})). In Fig.~\ref{fig:throughput}, we have that $N_4 = 64$, $N_3 = 128$, $N_2 = 192$, and $N_1 = 256$ for both IR and CC schemes. In addition, we consider another IR-HARQ scheme based on the proposed RCPP code to  send $k=32$ information bits via five different rates as
\begin{equation*}
r_1=\frac{32}{256} < r_2=\frac{32}{192} < r_3=\frac{32}{144} < r_4=\frac{32}{96} < r_5 = \frac{32}{64}.
\end{equation*} We remark that, in this example, the lengths of the (punctured) polar codes in the family are less likely to be the form of power of 2.  Similarly to the previous example, we observe that the proposed RCPP code can outperform the benchmark scheme due to the use of {\em improved} information sets effectively at the decoding side. Obviously, we can expect that, when a RCPP code should support a wide range of rates, the performance gain of the proposed method becomes much larger. 

\begin{figure}
\centerline{\includegraphics[width=10cm]{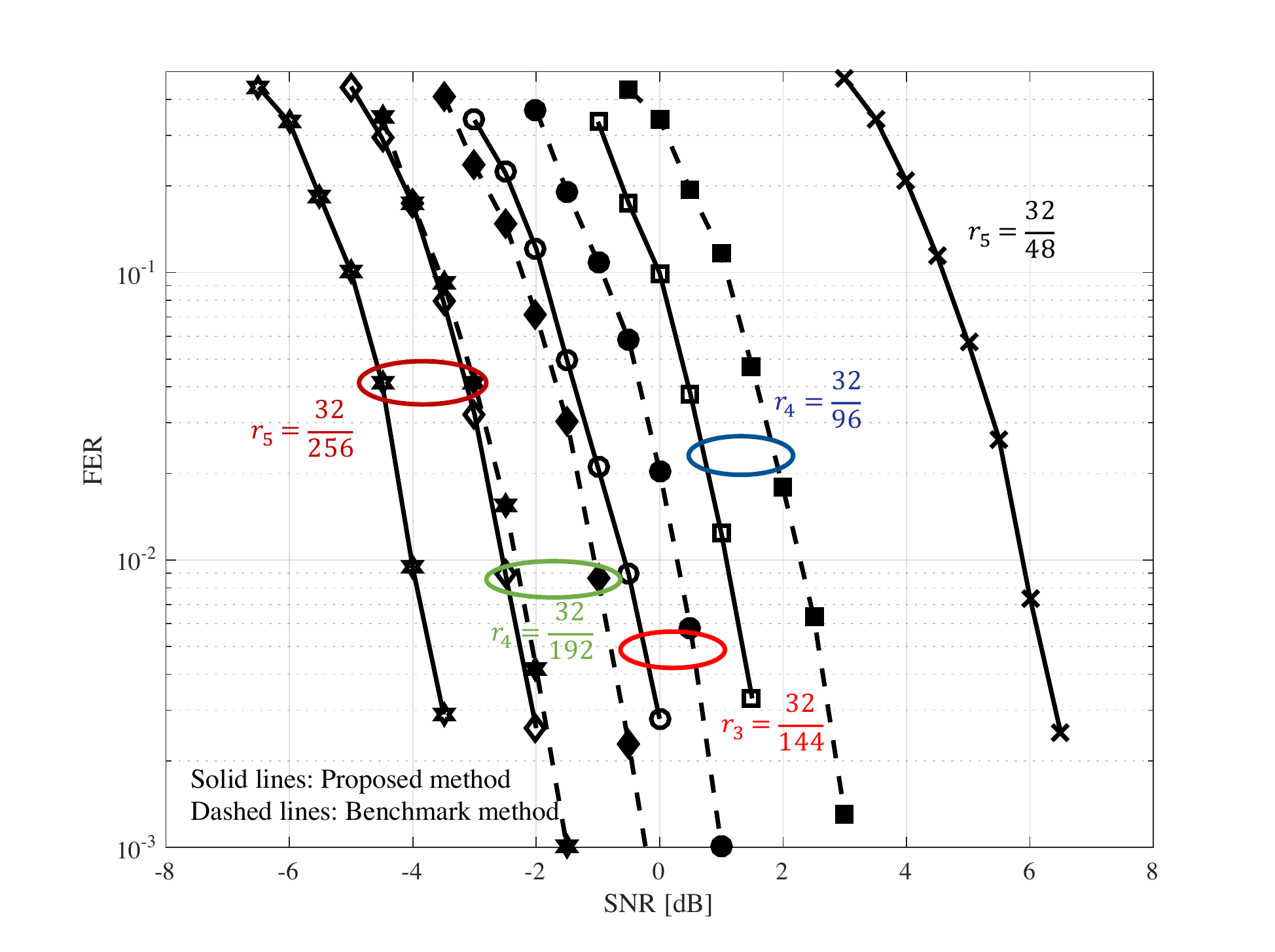}}
\caption{Performance comparison of the proposed and benchmark methods to support various rates $r_1 = \frac{32}{256} < r_2 = \frac{32}{192} < r_3 = \frac{32}{144} < r_4 = \frac{32}{96} < r_5 = \frac{32}{48}$. Both schemes employ the identical rate-compatible puncturing patterns (known as QUP) and the common information set $\Ac$ (which is optimized for the highest-rate code). The only difference is that the proposed approach uses the information-dependent frozen vector while the benchmark scheme uses the conventional all-zero frozen vector.}
\label{fig:sim2}
\end{figure}


\section{Conclusion}\label{sec:conc}

We presented novel reciprocal and hierarchical puncturing patterns, and derived their special properties. Leveraging them and the so-called {\em information-copy} technique, it is enabled that each code in the family of the proposed RCPP code can be decoded using its own optimized information set. Thus, the proposed method can address a challenging problem that, in conventional RCPP codes, one common information set was employed for all the codes in the family, thus not achieving balanced performances. Via simulation results, we demonstrated that the proposed RCPP code provides a non-trivial performance gain mainly due to the use of {\em enhanced} effective information sets. Therefore, it would be an important technique to construct a good RCPP code. An interesting future work is to identify a good hierarchical puncturing pattern for a given IR-HARQ system.

\end{document}